\documentclass[10pt]{article}

\usepackage[utf8]{inputenc}
\usepackage[letterpaper, top = 1.25in, bottom = 1.25in, left = 1.5in, right = 1.5in]{geometry}

\usepackage{titlesec}
\titleformat{\section}[hang]{\large\bfseries}{\thesection}{0.5cm}{}
\titleformat{\subsection}[hang]{\normalsize\bfseries}{\thesubsection}{0.3cm}{}

\usepackage{booktabs}
\usepackage{multirow}
\usepackage{graphicx}
\graphicspath{ {Figures/} }
\usepackage{subcaption}

\usepackage{mathtools, amssymb, bm, amsthm}
\newtheorem*{assumption}{Assumption}

\usepackage{tikz}
\usepackage{tkz-euclide}
\usetikzlibrary{shapes, arrows, positioning, calc, intersections}
\tikzstyle{block} = [draw, thick, rectangle, fill=white!20, minimum height=3em, minimum width=6em]
\tikzstyle{square} = [draw, thick, rectangle, fill=white!20, minimum height = 3em, minimum width = 3em]
\tikzstyle{coord} = [coordinate]
\tikzstyle{sum} = [draw, thick, circle, node distance = 1.5cm]

\usepackage{cite}
\usepackage[hidelinks]{hyperref}

\newcommand{\define}{ \stackrel{\Delta}{=} }
\newcommand{\lra}{ \longrightarrow }

\allowdisplaybreaks

\title{\Large{\textbf{Discrete Robust Control of Robot Manipulators\\using an Uncertainty and Disturbance Estimator}}}
\author{Ram Padmanabhan\thanks{Department of Electrical Engineering and Computer Science, University of Michigan, Ann Arbor, MI 48105, USA.}, Maithili Shetty\footnotemark[1], and T. S. Chandar\thanks{Department of Electronics and Communication Engineering, PES University, Ring Road Campus, Bengaluru 560085, India. Email: \texttt{chandarts@pes.edu}}}
\date{}

\begin{document}

\maketitle

\begin{abstract}
This article presents the design of a robust observer based on the discrete-time formulation of Uncertainty and Disturbance Estimator (UDE), a well-known robust control technique, for the purpose of controlling robot manipulators. The design results in a complete closed-loop, robust, controller--observer structure. The observer incorporates the estimate of the overall uncertainty associated with the plant, in order to mimic its dynamics, and the control law is generated using an auxiliary error instead of state tracking error. A detailed qualitative and quantitative stability analysis is provided, and simulations are performed on the two-link robot manipulator system. Further, a comparative study with well-known control strategies for robot manipulators is presented. The results demonstrate the efficacy of the proposed technique, with better tracking performance and lower control energy compared to other strategies.
\end{abstract}

\section{Introduction} \label{sec:Introduction}
Robot manipulators are widely used in various industrial applications, and have rapidly progressed from simple `pick-and-place' robots, to modern robots performing sophisticated activities in semiconductor manufacturing and medicine. The design of control strategies for manipulating robots to perform various tasks is a fundamental topic in robotics. Numerous strategies have been proposed for this, and can be found in the survey paper \cite{Survey1} as well as the books \cite{Spong, Murray}. Complications in this design arise due to the highly nonlinear dynamics of robot manipulators, and modelling errors and disturbances that are common in any engineering system. Simple approaches based on feedback linearization \cite{Spong, Khalil} are likely to fail due to this, and hence the robust control of robot manipulators is a highly active area of research.

Many popular control strategies have been applied for robust control of robot manipulators, including Sliding Mode Control (SMC) \cite{SMC-RM, SMC-RM1, SMC-RM2, SMC-RM3, SMC-RM4}, disturbance observer \cite{DO1, DO2, DO3, DO4}, $\mathcal{H}_{\infty}$ control \cite{Hinf-1, Hinf-2}, time-delay control (TDC) \cite{TDC1, TDC2, TDC3}, extended state observer \cite{ESO1, ESO2}, neural networks \cite{NN-1, NN-2, NN-3, NNBook} and optimal control \cite{Opt1, Opt2}. Certain common drawbacks are present in most of these strategies. Bounds on overall system uncertainty are required to be known, and states are required for feedback. While position states can be measured, the velocities of joints in robot manipulators are not easily measured and fed back to the controller. Further, the design of an observer is complicated by system uncertainties, hence requiring robust state estimation.

The technique of Uncertainty and Disturbance Estimator (UDE) was proposed by Zhong and Rees in \cite{CT-UDE}, for continuous-time systems. This is a simple frequency-domain technique, that estimates and compensates the overall uncertainty that a system is subjected to. The uncertainty estimate is modeled as a low-pass filter acting on the overall `lumped' uncertainty, and is compensated through the control law. Unlike SMC and other techniques, UDE does not require knowledge of bounds on overall system uncertainty, and unlike TDC, does not require system state derivatives for control design. The design of a robust observer for UDE has also been investigated \cite{UDE-Obs, CT-UDE-RM}. In \cite{CT-UDE-RM}, the control of a two-link robot manipulator based on UDE was explored, along with the design of a robust observer to ensure that link velocities were not required for feedback in control design.

All techniques mentioned thus far are continuous-time in nature. However, strategies based on discrete-time control laws are easier to implement on a microcontroller, by sampling system states or error signals and using a zero-order-hold to drive the original continuous-time system. This framework has also led to widespread use in the control of robot manipulators. In \cite{DT-SMC-RM}, standard discrete-time SMC reaching laws were used for finite-time control of robot manipulators. Tsai \emph{et al.} \cite{RC-RM} presented a discrete-time repetitive control strategy in combination with a proportional-derivative law for tracking periodic trajectories. An optimal radial-basis function neural network (RBFNN) was used in \cite{DT-NN-RM} for feedback control of robot manipulators. In \cite{AT-RM}, a discrete-time acceleration/torque controller was used to compute position, velocity and acceleration bounds in robot manipulator joints, and Kali \emph{et al.} designed a discrete second-order sliding mode in conjunction with time-delay estimation for trajectory tracking control of robotic arms \cite{DT-TDC-RM}.

In a recent work \cite{DT-UDE}, the authors proposed a discrete-time formulation for UDE, i.e. DT-UDE. This involved the design of a novel digital filter for modeling the uncertainty and disturbance estimate, derivation of the control law and conditions for stability of the closed-loop system. Further, the control of continuous-time linear as well as nonlinear systems was achieved using sampling and a zero-order-hold. It was also indicated that the DT-UDE strategy is less sensitive to initial values of tracking error compared to continuous-time UDE, requiring lower control energy. The design of DT-UDE also enables implementation on a microcontroller or digital computer. A drawback of the design in \cite{DT-UDE} is the requirement of all states for feedback for control design. This is not realistic in common engineering systems, as mentioned earlier, and is addressed here.

The objective of this work is to present a discrete controller for robot manipulators based on the formulation of DT-UDE in \cite{DT-UDE}. The principal contributions are summarized as follows. A robust observer based on the DT-UDE framework is designed. The observer uses the disturbance estimate in its dynamics, in order to mimic the plant. For a robot manipulator system, this design ensures that link velocities are not required to be fed back for control design, instead using only link positions which act as system outputs. Further, the design of the UDE-based control law uses state estimates to form an auxiliary tracking error, instead of using state tracking error. This results in the design of a DT-UDE-based controller--observer structure. A qualitative and quantitative analysis of stability is presented, and conditions on the parameter of the digital filter that models the disturbance estimate are derived. The quantitative analysis presents a new result on the convergence of the overall closed-loop error norm, based on the degree that the disturbance varies with time. The entire strategy is simulated for the two-link robot manipulator system, and the results indicate highly accurate tracking performance, with excellent disturbance rejection. Finally, comparative studies are presented, which demonstrate that the DT-UDE-based controller--observer structure outperforms well-known control strategies for robot manipulators, in terms of improved tracking performance and lower utilization of control energy.

The remainder of this article is organized as follows. Section \ref{sec:RMDyn} provides an overview of the dynamics of the two-link robot manipulator system, subsequently written in the state-space formulation. The tracking objective is also presented here. Section \ref{sec:UDEReview} reviews the theory of DT-UDE according to \cite{DT-UDE}. In Section \ref{sec:Design}, a discrete-time observer based on DT-UDE is designed, along with the control law, resulting in a robust controller--observer structure. A detailed stability analysis is provided in Section \ref{sec:Stability}. Section \ref{sec:Sim1} presents simulation results for the controller--observer structure applied to the two-link robot manipulator problem, and comparative studies are presented in Section \ref{sec:Sim2}. Concluding remarks are provided in Section \ref{sec:Conclusion}.

Throughout this article, $\mathbb{R}^p$ denotes the set of all $p$-tuples of real numbers, and $\mathbb{R}^{p\times q}$ denotes the set of all matrices with $p$ rows and $q$ columns, with real entries. $I_n$ denotes the $n\times n$ identity matrix, and $\bm{0}$ denotes the $n\times n$ matrix of zeros. The Moore-Penrose pseudo-inverse of a matrix $B \in \mathbb{R}^{m\times n}$ with full column rank is a matrix $B^{\dagger} \in \mathbb{R}^{n\times m}$, defined as $B^{\dagger} = \left(B^TB\right)^{-1}B^T$ and $B^{\dagger}B = I_n$. $\rho(A)$ denotes the spectral radius of the square matrix $A \in \mathbb{R}^{n\times n}$, defined as $\rho(A) = \max_{1\leq i \leq n} \left|\lambda_i(A)\right|$, where $\lambda_i(A)$ is the $i$th eigenvalue of $A$. A matrix $A$ is said to be a Schur matrix if $\rho(A) < 1$. For a symmetric, positive definite matrix $P$, the quantity $x^TPx$ is bounded as follows:
$$
p_{\mathrm{min}}\|x\|^2 \leq x^TPx \leq p_{\mathrm{max}}\|x\|^2,
$$
where $p_{\mathrm{min}}$ and $p_{\mathrm{max}}$ denote the minimum and maximum eigenvalues (both positive) of $P$. Further, $\|P\| = p_{\mathrm{max}}$.

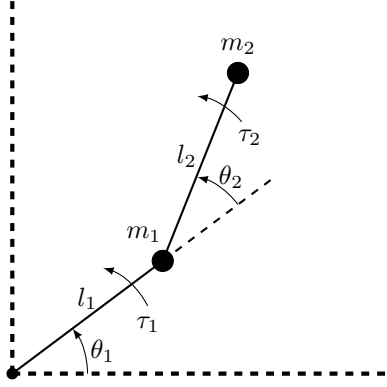
\begin{figure}
\begin{center}
\begin{tikzpicture}[auto,>=latex',connect with angle/.style=
{to path={let \p1=(\tikztostart), 
\p2=(\tikztotarget) in -- ++({\x2-\x1-(\y1-\y2)*tan(#1-90)},0) -- (\tikztotarget)}}]

\node[coord] at (2, 1.5) (m1) {};
\node[coord] at (3, 4) (m2) {};

\draw[ultra thick, dashed] (0, 0) -- (5, 0);
\draw[ultra thick, dashed] (0, 0) -- (0, 5);
\draw[thick] (0, 0) -- (m1) node[xshift = -1cm, yshift = -0.5cm]{$l_1$};
\draw[thick] (m1) -- (m2) node[xshift = -0.68cm, yshift = -1.1cm]{$l_2$};
\draw[thick, dashed] (m1) -- (3.5, 2.625);
\filldraw[black] (0, 0) circle(2pt) node{};
\filldraw[black] (m1) circle (4pt) node[xshift = -0.25cm, yshift = 0.35cm]{$m_1$};
\filldraw[black] (m2) circle (4pt) node[yshift = 0.35cm]{$m_2$};
\draw[-latex](1, 0) arc (0:36.87:1) node[xshift = 0.4cm, yshift = -0.3cm]{$\theta_1$};
\draw[-latex](3, 2.25) arc (36.87:75:1) node[xshift = 0.45cm, yshift = 0cm]{$\theta_2$};
\draw[-latex](1.8, 0.9) arc (26.57:72.5:1) node[xshift = 0.6cm, yshift = -0.75cm]{$\tau_1$};
\draw[-latex](3.05, 3.35) arc (40:80:1) node[xshift = 0.7cm, yshift = -0.55cm]{$\tau_2$};

\end{tikzpicture}
\end{center}
\caption{Schematic diagram of a two-link robot manipulator.}
\label{fig:RM}
\hrulefill
\end{figure}

\section{Dynamics of a Two-link Robot Manipulator} \label{sec:RMDyn}
A simple schematic diagram of the two-link robot manipulator is shown in Fig.~\ref{fig:RM}. Via the Euler-Lagrange formalism, the following well-known dynamics for this system can be obtained \cite{Nijmeijer}:

\begin{equation} \label{eq:RMDyn1}
	M(\theta)\Ddot{\theta} + C(\theta, \dot{\theta}) + K(\theta) = \tau,
\end{equation}
where
\begin{subequations} \label{eq:Matrices}
\begin{equation}
M(\theta) = \begin{bmatrix} m_1l_{1}^{2} + m_2\left|l_1 + l_2\right|^2 & m_2\left(l_{2}^{2} + l_1l_2\cos\theta_2\right) \\ m_2\left(l_{2}^{2} + l_1l_2\cos\theta_2\right) & m_2l_{2}^{2} \end{bmatrix}
\end{equation}
\begin{equation}
C(\theta, \dot{\theta}) = \begin{bmatrix} -m_2l_1l_2\left(\sin\theta_2\right)\dot{\theta}_2\left(2\dot{\theta}_1+\dot{\theta}_2\right) \\ m_2l_1l_2\left(\sin\theta_2\right)\dot{\theta}_{1}^{2} \end{bmatrix}
\end{equation}
\begin{equation}
K(\theta) = \begin{bmatrix} \left(m_1+m_2\right)gl_1\sin\theta_1 + m_2gl_2\sin\left(\theta_1 + \theta_2\right) \\ m_2gl_2\sin\left(\theta_1 + \theta_2\right) \end{bmatrix}.
\end{equation}
\end{subequations}

Here, $\theta = \left[\theta_1, \theta_2\right]^T$, $\dot{\theta}$ and $\Ddot{\theta}$ denote the two-dimensional vectors of link positions, velocities and accelerations respectively, and $\tau = \left[\tau_1, \tau_2\right]^T$ denotes the two-dimensional input torque vector. $M(\theta)$ denotes an inertia matrix, $C(\theta, \dot{\theta})$ reflects the centripetal and Coriolis forces and $K(\theta)$ characterizes the gravitational forces. $l_1$ and $l_2$ denote the lengths of the two links, $m_1$ and $m_2$ are the masses shown in the figure, and $g$ is the gravitational acceleration. In $M(\theta)$, we use $\left|l_1 + l_2\right|^2 = l_{1}^{2} + l_{2}^{2} + 2l_1l_2\cos\theta_2$. The matrix $M(\theta)$ is positive definite for all values of $\theta$, and hence the above dynamics \eqref{eq:RMDyn1} can be re-written as:

\begin{equation} \label{eq:RMDyn2}
	\Ddot{\theta} = -M(\theta)^{-1}C(\theta, \dot{\theta}) - M(\theta)^{-1}K(\theta) + M(\theta)^{-1}\tau.
\end{equation}

Let $M(\theta) = M_0 + \Delta M(\theta)$, where $M_0$ is a known, constant diagonal matrix and $\Delta M(\theta)$ denotes modelling uncertainties and other nonlinearities. Further, let $d'$ denote any additive external disturbances that affect the system \eqref{eq:RMDyn2}. Then,

\begin{equation} \label{eq:RMDyn3}
	\Ddot{\theta} = -M(\theta)^{-1}\big[C(\theta, \dot{\theta}) + K(\theta)\big] + \left(M(\theta)^{-1} - M_{0}^{-1}\right)\tau + M_{0}^{-1}\tau + d'.
\end{equation}
Define $d \define d' - M(\theta)^{-1}\big[C(\theta, \dot{\theta}) + K(\theta)\big] + \left(M(\theta)^{-1} - M_{0}^{-1}\right)\tau$. $d$ denotes the total disturbance acting on the system, and incorporates the matrices $M(\theta)$, $C(\theta, \dot{\theta})$ and $K(\theta)$ to mitigate modeling errors that may be present in these quantities. Then,

\begin{equation} \label{eq:RMDyn4}
	\Ddot{\theta} = d + M_{0}^{-1}\tau.
\end{equation}
Define the state vector $x \define \left[\theta_1, \dot{\theta}_1, \theta_2, \dot{\theta}_2\right]^T$. From \eqref{eq:RMDyn4},
\begin{align*}
	\dot{x}_1 &= x_2 \\
	\dot{x}_2 &= d_1 + \mu_1\tau_1 \\
	\dot{x}_3 &= x_4 \\
	\dot{x}_4 &= d_2 + \mu_2\tau_2,
\end{align*}
where $d_1$ and $d_2$ are the components of the total disturbance $d$. $\mu_1$ and $\mu_2$ denote the diagonal elements of $M_{0}^{-1}$, and are generally taken as the inverse of the diagonal elements of the inertia matrix $M(\theta)$, with the cosine terms set to unity. Explicitly indicating the dependence on time $t$, the above equations reduce to:
\begin{equation} \label{eq:RMDynSSMat}
\dot{x}(t) = Ax(t) + Bu(t) + D(x, t),
\end{equation}
where
$$
A = \begin{bmatrix} 0 & 1 & 0 & 0 \\ 0 & 0 & 0 & 0 \\ 0 & 0 & 0 & 1 \\ 0 & 0 & 0 & 0 \end{bmatrix}, \hspace{0.1cm} B = \begin{bmatrix} 0 & 0 \\ \mu_1 & 0 \\ 0 & 0 \\ 0 & \mu_2 \end{bmatrix}, \hspace{0.1cm} D(x, t) = \begin{bmatrix} 0 \\ d_1(t) \\ 0 \\ d_2(t) \end{bmatrix},
$$
and $u = \left[\tau_1(t), \tau_2(t)\right]^T$. $D(x, t)$ denotes the overall disturbance associated with the system, consisting of modelling uncertainties, nonlinearities and external disturbances. The two quantities available for feedback are the two link positions. Hence, define the output vector:
\begin{equation} \label{eq:RMDynOP}
	y(t) = \begin{bmatrix} \theta_1(t) \\ \theta_2(t) \end{bmatrix} = \begin{bmatrix} 1 & 0 & 0 & 0 \\ 0 & 0 & 1 & 0 \end{bmatrix} \begin{bmatrix} \theta_1(t) \\ \dot{\theta}_1(t) \\ \theta_2(t) \\ \dot{\theta}_2(t) \end{bmatrix} = Cx(t).
\end{equation}
Equations \eqref{eq:RMDynSSMat} and \eqref{eq:RMDynOP} describe the two-link robot manipulator system in state-space form. The tracking objective is to ensure that the states $x(t)$ in \eqref{eq:RMDynSSMat} follow the states of a stable reference model, described by:
\begin{equation} \label{eq:CTRefModel}
\dot{x}_m(t) = A_mx_m(t) + B_mr(t),
\end{equation}
where the dimensions of $x_m(t)$, $A_m$, $B_m$ and $r(t)$ match those of $x(t)$, $A$, $B$ and $u(t)$. In the following section, a brief review of discrete-time UDE, or DT-UDE is provided. The multiple-input, multiple-output case, which was not dealt with in detail in \cite{DT-UDE}, is considered here. The DT-UDE strategy, along with the design of a robust controller--observer structure in Section \ref{sec:Observer} enable the design of a control law that is robust to modeling uncertainties and any external disturbances, without requiring the measurement of link velocities.

\section{Review of Discrete-time Uncertainty and Disturbance Estimator} \label{sec:UDEReview}
Consider a discrete-time (DT), linear, time-invariant, multiple-input, multiple-output system with $n$ states and $p$ inputs:
\begin{align}
	x(k+1) &= \left(F_n + \Delta F_n\right)x(k) + \left(G_n + \Delta G_n\right)u(k) + d_d(x, k) \nonumber \\
	&= F_nx(k) + G_nu(k) + D_d(x, k). \label{eq:Plant}
\end{align}
$x(k) = \big[x_1(k), \ldots, x_n(k)\big]^T \in \mathbb{R}^n$ is the state vector and $u(k) = \big[u_1(k), \ldots, u_p(k)\big]^T \in \mathbb{R}^p$ is the input vector to be designed. $F_n \in \mathbb{R}^{n\times n}$ and $G_n \in \mathbb{R}^{n\times p}$ describe the `nominal' or `known' parameters of the plant, and $\Delta F_n \in \mathbb{R}^{n\times n}$ and $\Delta G_n \in \mathbb{R}^{n\times p}$ are their associated uncertainties. $d_d(x, k) \in \mathbb{R}^n$ is an external, possibly time-varying vector of disturbances. $D_d(x, k) \in \mathbb{R}^n$ is the overall disturbance associated with the plant, and is defined as:
\begin{equation}
	D_d(x, k) = \Delta F_nx(k) + \Delta G_nu(k) + d_d(x, k).
\end{equation}
The pair $\big\{F_n, G_n\big\}$ is assumed to be controllable, and sampling time of the system is $T_s$.

The objective is to ensure that states of the system \eqref{eq:Plant} follow the states of a stable discrete-time reference model:
\begin{equation} \label{eq:RefModel}
	x_m(k+1) = F_mx_m(k) + G_mr(k).
\end{equation}
$x_m(k) = \big[x_{1m}(k), \cdots, x_{nm}(k)\big]^T \in \mathbb{R}^n$ is the state vector of the reference model, $F_m \in \mathbb{R}^{n\times n}$ and $G_m \in \mathbb{R}^{n\times p}$ are matrices describing the reference model and $r(k) \in \mathbb{R}^p$ is the input to the reference model. Note that the dimensions of reference model parameters mimic those of the plant.

Assuming all states are available for feedback, define the state tracking error $e(k) \define x(k) - x_m(k) = \big[e_1(k), \ldots, e_n(k)\big]^T \in \mathbb{R}^n$. Then,
\begin{equation} \label{eq:Err1}
	e(k+1) = F_ne(k) + G_nu(k) + L_d(x, k).
\end{equation}
$L_d(x, k) = D_d(x, k) + \big(F_n - F_m\big)x_m(k) - G_mr(k) \in \mathbb{R}^n$ denotes the `lumped' disturbance associated with the above dynamics. To simplify notation this is henceforth denoted as $L_d(k)$, but it must be kept in mind that the lumped disturbance depends on the state $x(k)$. If this quantity were zero, then an error feedback law $u(k) = -K_de(k)$ could be designed to stabilize the above system. However, for $L_d(k) \neq 0$, the following `matching condition' assumption is required.
\begin{assumption} \label{asm:Matching}
	The lumped disturbance $L_d(k)$ satisfies:
	\begin{equation} \label{eq:Matching}
		L_d(k) = G_n\psi_d(k),
	\end{equation}
	where $\psi_d(k) \in \mathbb{R}^{p}$ is some vector of disturbances.
\end{assumption}

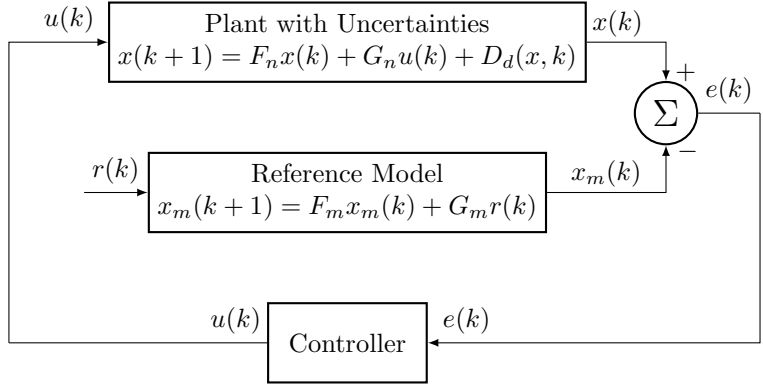
\begin{figure*}[t!]
\begin{center}
\begin{tikzpicture}[auto,>=latex',connect with angle/.style=
{to path={let \p1=(\tikztostart), 
\p2=(\tikztotarget) in -- ++({\x2-\x1-(\y1-\y2)*tan(#1-90)},0) -- (\tikztotarget)}}]

    \node [block, align = center] (plant) {Plant with Uncertainties\\$x(k+1) = F_nx(k) + G_nu(k) + D_d(x, k)$};
    \node [sum, below right = 0.1cm and 0.75cm of plant] (Sum) {\Large$\Sigma$};
    \node [block, below of = plant, node distance = 2cm, align = center] (refmod) {Reference Model\\$x_m(k+1) = F_mx_m(k) + G_mr(k)$};
    \node [block, below of = refmod, node distance = 2cm] (controller) {Controller};
    \node [coord, left of = refmod, node distance = 3.5cm] (ref) {};
    \node [coord, right of = Sum, node distance = 1.25cm] (inter1) {};
    \node [coord, left of = controller, node distance = 4.5cm] (inter2) {};

    \draw[-latex] (ref) -- node{$r(k)$} (refmod);
    \draw[-latex] (plant) -| node[xshift = -1.1cm, yshift = 0.25cm]{$x(k)$} node[yshift = -0.4cm] {$+$} (Sum);
    \draw[-latex] (refmod) -| node[xshift = -0.2cm, yshift = 0.25cm]{$x_m(k)$} node[xshift = 0.55cm, yshift = 0.55cm] {$-$} (Sum);
    \draw[-] (Sum) -- node{$e(k)$} (inter1);
    \draw[-latex] (inter1) |- node[xshift = -3.9cm, yshift = 0.6cm]{$e(k)$} (controller);
    \draw[-] (controller) -- node[xshift = 1.3cm, yshift = 0.6cm]{$u(k)$} (inter2);
    \draw[-latex] (inter2) |- node[xshift = 0.8cm, yshift = 0cm]{$u(k)$} (plant);

\end{tikzpicture}
\end{center}
\caption{Block diagram for simple discrete-time UDE-based controller.}
\label{fig:DT-UDE}
\hrulefill
\end{figure*}

Subsequently, construct an estimate of the disturbance $\hat{L}_d(k) = G(\gamma)L_d(k)$, where $\displaystyle G(\gamma) = \frac{1}{1 + \tau\gamma}$, $\gamma = (z-1)\big/T_s$, and $z$ is the digital frequency-domain operator. $\tau$ is simply a parameter of the filter. Note that the notation $G(\gamma)$ is not to be confused with the input matrices $G_n$ or $G_m$. Design the control input:
\begin{equation} \label{eq:U1}
	u(k) = -K_de(k) - G_{n}^{\dagger}\hat{L}_d(k) = -K_de(k) + u_d(k),
\end{equation}
where $K_d \in \mathbb{R}^{p\times n}$ is designed such that $\rho(F_n - G_nK_d) < 1$, and $G_nu_d(k) = -\hat{L}_d(k)$. This control law seeks to nullify the effect of $L_d(k)$ by subtracting its estimate from the error system \eqref{eq:Err1}. Using \eqref{eq:U1} in \eqref{eq:Err1}:
\begin{equation}
	e(k+1) = (F_n - G_nK_d)e(k) + G_nu_d(k) + L_d(k) \label{eq:Err2},
\end{equation}
or,
$$
	L_d(k) = e(k+1) - (F_n - G_nK_d)e(k) - G_nu_d(k).
$$
Multiplying throughout by $G_{n}^{\dagger}G(\gamma)$, and using the fact that $G_{n}^{\dagger}G(\gamma)L_d(k) = G_{n}^{\dagger}\hat{L}_d(k) = -u_d(k)$:
$$
	-u_d(k) = G(\gamma)G_{n}^{\dagger}\Big(e(k+1) - \big(F_n - G_nK_d\big)e(k)\Big) - G(\gamma)u_d(k).
$$
Solving this for $u_d(k)$:
$$
u_d(k) = -\frac{T_s}{\tau}\sum_{n = 0}^{k}G_{n}^{\dagger}\Big(e(n) - \big(F_n - G_nK_d\big)e(n-1)\Big).
$$
The above expression is obtained on replacing the quantity $G(\gamma)\big/(1 - G(\gamma)) = 1\big/(1-z^{-1})$ by its action, an accumulator/running sum. For $k = 0$ however, this expression is not valid. Taking this into account,
\begin{equation} \label{eq:UD}
	u_d(k) = \begin{dcases} -\frac{T_s}{\tau}G_{n}^{\dagger}e(0) &\text{if } k = 0 \\ -\frac{T_s}{\tau}\sum_{n = 1}^{k}G_{n}^{\dagger}\Big(e(n) - \big(F_n - G_nK_d\big)e(n-1)\Big) \hspace{0.5cm} &\text{if } k = 1, 2, \ldots
	\end{dcases}
\end{equation}
Hence, the overall control law from \eqref{eq:U1} and \eqref{eq:UD} is:
\begin{equation} \label{eq:Control}
	u(k) = \begin{dcases} -\left(K_d + \frac{T_s}{\tau}G_{n}^{\dagger}\right)e(0) &\text{if } k = 0 \\ -K_de(k) -\frac{T_s}{\tau}\sum_{n = 1}^{k}G_{n}^{\dagger}\Big(e(n) - \big(F_n - G_nK_d\big)e(n-1)\Big) \hspace{0.5cm} &\text{if } k = 1, 2, \ldots \end{dcases}
\end{equation}
Figure \ref{fig:DT-UDE} shows a simple and general block diagram for the procedure outlined above, with $e(k)$ being used as in eq. \eqref{eq:Control} to generate $u(k)$.

A few general remarks on the discrete-time UDE (DT-UDE) strategy described above are in order. One of the most important aspects to analyze in the closed-loop system described by the plant \eqref{eq:Plant}, reference model \eqref{eq:RefModel} and the control law \eqref{eq:Control} is stability. While not reproduced here, it has been shown in \cite{DT-UDE} that the condition on the filter parameter $\tau$ to ensure asymptotic stability of the system is:
\begin{equation} \label{eq:Stability}
	\tau > \frac{T_s}{2},
\end{equation}
provided that the quantity $\Delta L_d(k) = L_d(k+1) - L_d(k) \approx 0$, i.e. the lumped disturbance is slowly varying. With fast-varying disturbances, only bounded-input, bounded-output (BIBO) stability is ensured. 

Next, the importance of DT-UDE arises from the fact that real-world control engineering systems incorporate embedded microcontrollers that process discrete-time, or digital data. A digital control law as in \eqref{eq:Control} can be easily generated by such controllers. However, physical, real-world systems are continuous-time systems of the form \eqref{eq:RMDynSSMat}, and are primarily nonlinear in nature. Continuous-time signals cannot be processed by digital computers and microcontrollers. To address this, a sampler is used to sample the state tracking error, the control law \eqref{eq:Control} is generated using this, and finally a zero-order-hold, i.e. a digital--analog converter is used to drive the original system. Further, nonlinearities are incorporated into the lumped disturbance, hence ensuring that only linear terms affected by a lumped disturbance remain as part of the system. This procedure has been explored in detail in \cite{DT-UDE}, with extensive simulations on both linear, time-invariant as well as nonlinear continuous-time systems.

Note that with a general continuous-time plant as in \eqref{eq:RMDynSSMat} and reference model in \eqref{eq:CTRefModel}, the dynamics of state tracking error can be obtained as:
$$
	\dot{e}(t) = Ae(t) + Bu(t) + L(t),
$$
where $L(t) = D(x, t) + (A - A_m)x_m(t) - B_mr(t)$ denotes the `lumped' disturbance for the tracking error dynamics. On sampling the state tracking error with sampling time $T_s$, discrete-time dynamics of the form \eqref{eq:Err1} are obtained, where:
$$
F_n = \exp{\left(AT_s\right)}; \hspace{0.25cm} G_n = \int_{\theta = 0}^{T_s}\exp{\left(A\theta\right)}B\mathrm{d}\theta; \hspace{0.25cm} L_d(k) = \int_{\theta = 0}^{T_s}\exp{(A\theta)}L\left( (k+1)T_s - \theta\right)\mathrm{d}\theta.
$$
Similarly, when the states of the continuous-time reference model $x_m(t)$ from \eqref{eq:CTRefModel} are sampled, the resulting discrete-time dynamics are of the form \eqref{eq:RefModel}, where:
$$
F_m = \exp{\left(A_mT_s\right)}; \hspace{0.5cm} G_m = \int_{\theta = 0}^{T_s}\exp{\left(A_m\theta\right)}B_m\mathrm{d}\theta.
$$

Finally, consider the application of DT-UDE to the two-link robot manipulator system described in Section \ref{sec:RMDyn}. Throughout the treatment of DT-UDE here, it is assumed that the entire state vector $x(k)$ is accessible, so that the state tracking error $e(k)$ and subsequently the control law $u(k)$ in \eqref{eq:Control} can be generated. In the two-link robot manipulator system in \eqref{eq:RMDynSSMat}, the link positions are measurable states, but the link velocities are not. To design this, an observer/estimator must be designed to ensure that the control law \eqref{eq:Control} can be designed by measuring only the system outputs in \eqref{eq:RMDynOP}, i.e. the link positions. A simple Luenberger observer will not result in satisfactory performance however, due to the presence of uncertainties and disturbances in the system. This is addressed in the following section.

Further, the system \eqref{eq:RMDynSSMat} is a continuous-time system. As described earlier, if states were available for feedback, then the state tracking error $e(t)$ would be sampled to generate $e(k)$, to be used in \eqref{eq:Control}. To drive the original system, a zero-order-hold would be used to generate a continuous-time law. However, not all states are available for feedback in the considered system. A different, auxiliary error that uses only state estimates must be generated and used in the control law. Further, since the observer is not a physical system, but rather a \emph{virtual} one, it must be implemented in discrete-time, though the original plant is continuous-time. The state estimates (and hence the auxiliary error signal) are all discrete-time quantities. This procedure is also described in the following section.

\section{Discrete-time Robust Controller--Observer Structure} \label{sec:Observer}
In this section, the design of a robust discrete-time controller--observer structure is investigated, to address situations in which not all system states may be available for feedback in DT-UDE. The task of an observer is to mimic the dynamics of the plant. Due to the presence of $D_d(x, k)$ in \eqref{eq:Plant}, the usual Luenberger observer cannot achieve satisfactory performance. However, note that an estimate of $L_d(k)$ is available as part of the control law $u(k)$, from the term $u_d(k)$. This fact is used to design an observer to mimic plant dynamics. Further, since plant states are unavailable, the state estimate $\hat{x}(k)$ from the observer is used to generate an auxiliary error signal $\hat{e}(k) = \hat{x}(k) - x_m(k)$, which mimics the state tracking error. This auxiliary error is used in the control law, instead of state tracking error. The UDE-based controller, in conjunction with the robust discrete-time observer represent a DT-UDE-based controller--observer structure.

\subsection{Observer Design and Control Law} \label{sec:Design}
Consider the same plant in\eqref{eq:Plant}, with an additional output equation:
\begin{subequations} \label{eq:PlantObs}
\begin{align}
	x(k+1) &= F_nx(k) + G_nu(k) + D_d(x, k) \\
	y(k) &= Cx(k).
\end{align}
\end{subequations}
$y(k) = \big[y_1(k), \ldots, y_q(k)\big]^T \in \mathbb{R}^q$ is the output vector, and $C \in \mathbb{R}^{q\times n}$ is the completely known output matrix. The remainder of the problem is formulated as earlier, with the same reference model structure \eqref{eq:RefModel} and assumption \eqref{eq:Matching}, with all quantities having the same meaning. In addition to the assumption of controllability on $\big\{F_n, G_n\big\}$, the pair $\big\{C, F_n\big\}$ is assumed to be observable, so that the realization $\big\{C, F_n, G_n\big\}$ is minimal.

Recall that $L_d(k) = D_d(x, k) + (F_n - F_m)x_m(k) - G_mr(k)$ in \eqref{eq:Err1}. Then, the estimate $\hat{L}_d(k)$ can be written as:
\begin{equation} \label{eq:DistEstRelation}
	\hat{L}_d(k) = \hat{D}_d(x, k) + (F_n - F_m)x_m(k) - G_mr(k),
\end{equation}
as the quantities $F_n$, $F_m$, $x_m(k)$, $G_m$ and $r(k)$ are all known and available. $\hat{D}_d(x, k)$ is the estimate of $D_d(x, k)$, and is generated from the above equation. $\hat{L}_d(k)$ is generated from $u_d(k)$, the robust control, as $\hat{L}_d(k) = -G_nu_d(k)$. Note that $\hat{D}_d(x, k)$ is not generated using $G(\gamma)$.

The plant in \eqref{eq:PlantObs} contains $D_d(x, k)$. To design an observer, the plant dynamics must be mimicked; however, $D_d(x, k)$ is a disturbance vector that is not available. Hence, its estimate $\hat{D}_d(x, k)$ is used in the observer dynamics, constructed as:
\begin{equation} \label{eq:Observer}
	\hat{x}(k+1) = F_n\hat{x}(k) + G_nu(k) + \hat{D}_d(x, k) + \beta\left(y(k) - \hat{y}(k)\right).
\end{equation}
This equation represents the simple Luenberger observer augmented with the estimate $\hat{D}_d(x, k)$. $\beta \in \mathbb{R}^{n\times q}$ is the observer gain matrix to be designed, and $\hat{y}(k) = C\hat{x}(k)$. Define the state estimation error $e_{SE}(k) = x(k) - \hat{x}(k) = \left[e_{{SE},1}(k), \cdots, e_{{SE},n}(k)\right]^T$. From eqs. \eqref{eq:PlantObs} and \eqref{eq:Observer}:
\begin{align*}
	e_{SE}(k+1) &= F_ne_{SE}(k) + D_d(x, k) - \hat{D}_d(x, k) - \beta Ce_{SE}(k) \\
	&= (F_n - \beta C)e_{SE}(k) + D_d(x, k) - \hat{D}_d(x, k).
\end{align*}
From \eqref{eq:DistEstRelation} and the expression for $L_d(k)$, $D_d(x, k) - \hat{D}_d(x, k) = L_d(k) - \hat{L}_d(k) = \Tilde{L}_d(k)$. Hence,
\begin{equation} \label{eq:StateEstErr}
	e_{SE}(k+1) = (F_n - \beta C)e_{SE}(k) + \Tilde{L}_d(k).
\end{equation}

Note that despite some progress in designing the observer \eqref{eq:Observer} and deriving dynamics for the state estimation error \eqref{eq:StateEstErr}, the control law still requires state tracking error, which can be formed only when all states are available for feedback. To this end, define an auxiliary error $\hat{e}(k) = \hat{x}(k) - x_m(k)$. This mimics the tracking error $e(k) = x(k) - x_m(k)$, and is used in its place in $u(k)$. From \eqref{eq:Observer} and \eqref{eq:RefModel},
\begin{align*}
	\hat{e}(k+1) &= F_n\hat{e}(k) + G_nu(k) + \hat{D}_d(x, k) + \beta Ce_{SE}(k) + (F_n - F_m)x_m(k) - G_mr(k).
\end{align*}
Finally, rewrite the control law by simply replacing the state tracking error $e(k)$ by the auxiliary error $\hat{e}(k)$. Generate $u(k) = -K_d\hat{e}(k) + u_d(k)$, designing $K_d$ such that $\rho(F_n - G_nK_d) < 1$, with $u_d(k)$ given by:
\begin{equation} \label{eq:UD-Observer}
	u_d(k) = \begin{dcases} -\frac{T_s}{\tau}G_{n}^{\dagger}\hat{e}(0) &\text{if } k = 0 \\ -\frac{T_s}{\tau}\sum_{n = 1}^{k}G_{n}^{\dagger}\Big(\hat{e}(n) - \big(F_n - G_nK_d\big)\hat{e}(n-1)\Big) \hspace{0.5cm} &\text{if } k = 1, 2, \ldots
	\end{dcases}
\end{equation}
Thus, $u(k)$ is given by:
\begin{equation} \label{eq:Control-Observer1}
	u(k) = \begin{dcases} -\left(K_d + \frac{T_s}{\tau}G_{n}^{\dagger}\right)\hat{e}(0) &\text{if } k = 0 \\ -K_d\hat{e}(k) -\frac{T_s}{\tau}\sum_{n = 1}^{k}G_{n}^{\dagger}\Big(\hat{e}(n) - \big(F_n - G_nK_d\big)\hat{e}(n-1)\Big) \hspace{0.5cm} &\text{if } k = 1, 2, \ldots \end{dcases}
\end{equation}
This is the input to both the plant and the observer. Finally, $\hat{e}(k)$ can be rewritten using $u(k) = -K_d\hat{e}(k) + u_d(k)$ and \eqref{eq:DistEstRelation} as:
\begin{equation} \label{eq:AuxErr}
    \hat{e}(k+1) = (F_n - G_nK_d)\hat{e}(k) + \beta Ce_{SE}(k).
\end{equation}
Substituting \eqref{eq:AuxErr} in \eqref{eq:Control-Observer1}, the simplified control law $u(k)$ is obtained:
\begin{equation} \label{eq:Control-Observer}
	u(k) = \begin{dcases} -\left(K_d + \frac{T_s}{\tau}G_{n}^{\dagger}\right)\hat{e}(0) &\text{if } k = 0 \\ 
	-K_d\hat{e}(k) -\frac{T_s}{\tau}\sum_{n = 1}^{k}G_{n}^{\dagger}\beta\big(y(k) - \hat{y}(k)\big) \hspace{0.5cm} &\text{if } k = 1, 2, \ldots \end{dcases}
\end{equation}
where the fact that $Ce_{SE}(k) = y(k) - \hat{y}(k)$ has been used.

A simple block diagram for the procedure in this section is shown in Fig.~\ref{fig:DT-UDE-Observer}. The contrast with Fig.~\ref{fig:DT-UDE} is clear: the controller cannot use the tracking error $e(k)$ in designing $u(k)$, as the plant states are not available. Instead, the auxiliary error $\hat{e}(k)$, mimicking $e(k)$, is used to design $u(k)$. 

\subsection{Stability Analysis} \label{sec:Stability}
In this section, the stability of the entire closed-loop controller--observer structure is analyzed. Given $\hat{L}_d(k) = G(\gamma)L_d(k)$ and $\Tilde{L}_d(k) = L_d(k) - \hat{L}_d(k)$, it is easily shown using a few algebraic manipulations (see \cite{DT-UDE}) that
\begin{equation} \label{eq:DisEstErr}
	\Tilde{L}_d(k+1) = \left(1 - \frac{T_s}{\tau}\right)\Tilde{L}_d(k) + \Delta L_d(k),
\end{equation}
where $\Delta L_d(k) = L_d(k+1) - L_d(k)$. Now, use \eqref{eq:StateEstErr} and \eqref{eq:AuxErr} to construct the following dynamics:
\begin{equation} \label{eq:ObsDyn}
\begin{bmatrix} \hat{e}(k+1) \\ e_{SE}(k+1) \\ \Tilde{L}_d(k+1) \end{bmatrix}
=
\begin{bmatrix} F_c & \beta C & \bm{0} \\ \bm{0} & F_o & I_n \\ \bm{0} & \bm{0} & T \end{bmatrix}
\begin{bmatrix} \hat{e}(k) \\ e_{SE}(k) \\ \Tilde{L}_d(k) \end{bmatrix} + 
\begin{bmatrix} \bm{0} \\ \bm{0} \\ I_n \end{bmatrix} \Delta L_d(k),
\end{equation}
where $F_c = F_n - G_nK_d$, $F_o = F_n - \beta C$ and $T = \left(1 - T_s/\tau\right)I_n$. It is interesting to note that the three main error signals involved in observer design --- the state estimation error, auxiliary error and disturbance estimation error --- are used to construct these dynamics. Further, the tracking error $e(k) = e_{SE}(k) + \hat{e}(k)$.

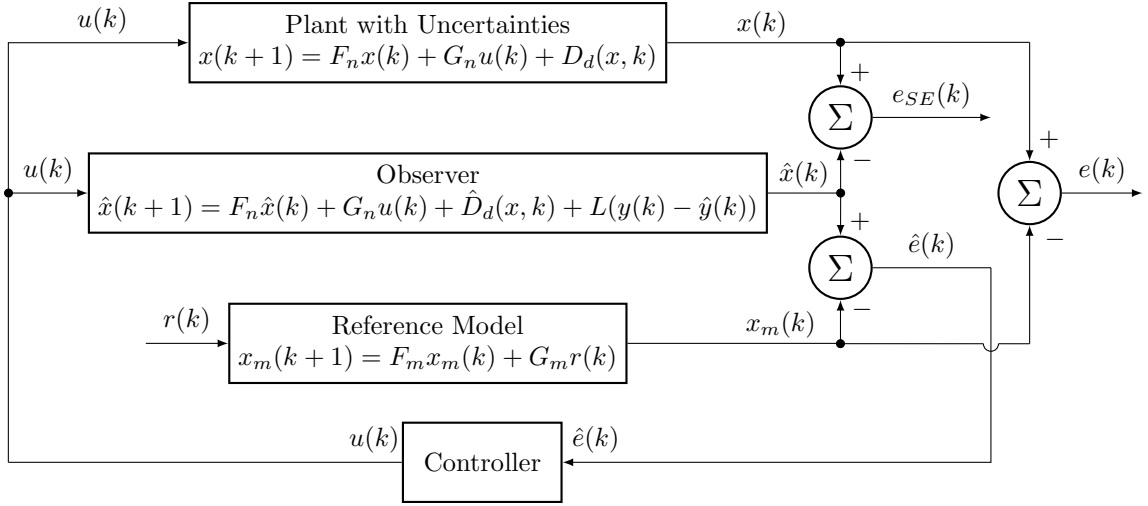
\begin{figure}[t!]
\begin{center}
\begin{tikzpicture}[auto,>=latex',connect with angle/.style=
{to path={let \p1=(\tikztostart), 
\p2=(\tikztotarget) in -- ++({\x2-\x1-(\y1-\y2)*tan(#1-90)},0) -- (\tikztotarget)}}]

    \node [block, align = center] (plant) {Plant with Uncertainties\\$x(k+1) = F_nx(k) + G_nu(k) + D_d(x, k)$};
    \node [sum, below right = 0.15cm and 2cm of plant] (Sum1) {\Large$\Sigma$};
    \node [block, below of = plant, node distance = 2cm, align = center] (observer) {Observer\\$\hat{x}(k+1) = F_n\hat{x}(k) + G_nu(k) + \hat{D}_d(x, k) + L(y(k) - \hat{y}(k))$};
    \node [coord, right of = observer, node distance = 5.4875cm] (interobs) {};
    \node [coord, left of = observer, node distance = 5.575cm] (obsinput) {};
    \node [sum, right of = observer, node distance = 8cm] (Sum3) {\Large$\Sigma$};
    \node [coord, right of = Sum3, node distance = 1.5cm] (tracking) {};
    \node [coord, above of = Sum3, node distance = 2.005cm] (interplant) {};
    \node [coord, below of = Sum3, node distance = 2.005cm] (interref) {};
    \node [coord, above of = Sum1, node distance = 1cm] (interplant1) {};
    \node [sum, below of = Sum1, node distance = 2cm] (Sum2) {\Large$\Sigma$};
    \node [coord, below of = Sum2, node distance = 1.01cm] (interref1) {};
    \node [coord, right of = Sum1, node distance = 2cm] (eSE) {};
    \node [coord, right of = Sum2, node distance = 2cm] (inter1) {};
    \node [block, below of = observer, node distance = 2cm, align = center] (refmod) {Reference Model\\$x_m(k+1) = F_mx_m(k) + G_mr(k)$};
    \node [block, below right = 0.5cm and -3cm of refmod] (controller) {Controller};
    \node [coord, left of = refmod, node distance = 3.75cm] (ref) {};
    \node [coord, left of = controller, node distance = 6.3cm] (inter2) {};

    \draw[-latex] (ref) -- node{$r(k)$} (refmod);
    \draw[-latex] (plant) -| node[xshift = -1.5cm, yshift = 0.25cm]{$x(k)$} node[yshift = -0.4cm] {$+$} (Sum1);
    \draw[-] (interplant1) -- node{} (interplant);
    \filldraw[black] (interplant1) circle (1.5pt) node{};
    \draw[-latex] (interplant) -- node[yshift = -0.5cm]{$+$} (Sum3);
    \path[name path = line2] (interref1) -- (interref);
    \filldraw[black] (interref1) circle (1.5pt) node{};
    \draw[-latex] (interref) -- node[xshift = 0.6cm, yshift = 0.6cm]{$-$} (Sum3);
    \draw[-latex] (Sum3) -- node{$e(k)$} (tracking);
    \draw[-latex] (refmod) -| node[xshift = -0.2cm, yshift = 0.25cm]{$x_m(k)$} node[xshift = 0.55cm, yshift = 0.45cm] {$-$} (Sum2);
    \draw[-] (observer) -- node[yshift = -0.05cm]{$\hat{x}(k)$} (interobs);
    \draw[-latex] (interobs) -- node[xshift = 0.55cm, yshift = 0.15cm]{$-$} (Sum1);
    \draw[-latex] (interobs) -- node[yshift = -0.1cm]{$+$} (Sum2);
    \filldraw[black] (interobs) circle (1.5pt) node{};
    \draw[-latex] (Sum1) -- node{$e_{SE}(k)$} (eSE);
    \draw[-] (Sum2) -- node{$\hat{e}(k)$} (inter1);
    \draw[-latex, name path = line1] (inter1) |- node[xshift = -5.25cm, yshift = 0.6cm]{$\hat{e}(k)$} (controller);
    \path [name intersections = {of = line1 and line2}];
    \coordinate (S)  at (intersection-1);
    \path[name path = circle] (S) circle(1mm);
    \path [name intersections={of = circle and line2}];
    \coordinate (I1)  at (intersection-1);
    \coordinate (I2)  at (intersection-2);
    \draw[-] (interref1) -- (I1);
    \draw[-] (I2) -- (interref);
    \tkzDrawArc[color=black](S,I1)(I2);
    \draw[-] (controller) -- node[xshift = 2.25cm, yshift = 0.6cm]{$u(k)$} (inter2);
    \draw[-latex] (inter2) |- node[xshift = 1.25cm, yshift = 0cm]{$u(k)$} (plant);
    \draw[-latex] (obsinput) -- node{$u(k)$} (observer);
    \filldraw[black] (obsinput) circle (1.5pt) node{};

\end{tikzpicture}
\end{center}
\caption{Block diagram for discrete-time UDE-based controller--observer structure.}
\label{fig:DT-UDE-Observer}
\hrulefill
\end{figure}

If $\Delta L_d(k)$ is sufficiently small, i.e. $L_d(k)$ is slowly time-varying, then the system \eqref{eq:ObsDyn} is asymptotically stable if the eigenvalues of the system matrix are within the unit disk on the $z$-plane, i.e.,
\begin{equation} \label{eq:ObsStabCond}
\rho(F_n - G_nK_d) < 1, \hspace{0.15cm} \rho(F_n - \beta C) < 1, \hspace{0.15cm} \left|1 - \frac{T_s}{\tau}\right| < 1.
\end{equation}
As the realization $\big\{C, F_n, G_n\big\}$ is minimal, $K_d$ and $\beta$ can be chosen such that the eigenvalues of $(F_n - G_nK_d)$ and $(F_n - \beta C)$ can be placed at desired locations within the unit disk on the $z$-plane. For satisfactory performance, the observer must respond much faster than the plant, and hence the eigenvalues of $(F_n - \beta C)$ must be much closer to the origin than those of $(F_n - G_nK_d)$. The third condition is then satisfied by ensuring:
\begin{equation} \label{eq:ObserverStab}
	\tau > \frac{T_s}{2}.
\end{equation}
This condition is the same as that obtained for the straightforward controller structure of Fig.~\ref{fig:DT-UDE} earlier \cite{DT-UDE}. With slowly varying $L_d(k)$, $\hat{e}(k) \lra 0$, $e_{SE}(k) \lra 0$ and $\Tilde{L}_d(k) \lra 0$ as $k \lra \infty$. Hence, the tracking error $e(k) = e_{SE}(k) + \hat{e}(k) \lra 0$ as $k \lra \infty$.

The entirety of the above analysis is qualitative in nature. It is also interesting to investigate the actual behaviour of each component in the dynamics \eqref{eq:ObsDyn} in a more quantitative manner. To this end, rewrite \eqref{eq:ObsDyn} in the following state-space formulation:
\begin{equation} \label{eq:ObsDynSS}
\xi(k+1) = \mathcal{A}\xi(k) + \eta(\xi(k), k).
\end{equation}
$\xi(k)$ denotes the state vector in \eqref{eq:ObsDyn}, containing the auxiliary error $\hat{e}(k)$, state estimation error $e_{SE}(k)$ and the disturbance estimation error $\Tilde{L}_d(k)$. $\eta(\xi(k), k)$ denotes the additional disturbance term depending on $\Delta L_d(k)$. For notational simplicity, $\xi(k)$ is denoted as $\xi_k$ and $\eta(\xi(k), k)$ is denoted as $\eta_k$ from this point onwards.

Note that $\mathcal{A}$ is a Schur matrix, as long as the conditions in \eqref{eq:ObsStabCond} are satisfied. Then, there exists a symmetric, positive definite matrix $P$ such that
\begin{equation} \label{eq:LyapunovDT}
\mathcal{A}^TP\mathcal{A} - P = -I_{3n}.
\end{equation}
Define a Lyapunov function
\begin{equation} \label{eq:LyapunovFn}
V_k \define  \xi_{k}^{T}P\xi_k.
\end{equation}
Then, $\Delta V_k = V_{k+1} - V_k$. Simplifying,
\begin{align}
\Delta V_k &= \left(\mathcal{A}\xi_k + \eta_k\right)^TP\left(\mathcal{A}\xi_k + \eta_k\right) - \xi_{k}^{T}P\xi_k \nonumber \\
&= \xi_{k}^{T}\left(\mathcal{A}^TP\mathcal{A} - P\right)\xi_k + 2\xi_{k}^{T}\mathcal{A}^TP\eta_k + \eta_{k}^{T}P\eta_k \nonumber \\
&= -\left\|\xi_k\right\|^2 + 2\xi_{k}^{T}\mathcal{A}^TP\eta_k + \eta_{k}^{T}P\eta_k. \label{eq:Lyapunov1}
\end{align}
As $P$ is a positive definite matrix, the inequality $p_{\mathrm{min}}\left\|\eta_k\right\|^2 \leq \eta_{k}^{T}P\eta_k \leq p_{\mathrm{max}}\left\|\eta_k\right\|^2$ holds, where $p_{\mathrm{min}}$ and $p_{\mathrm{max}}$ denote the minimum and maximum eigenvalues (both positive) of $P$ respectively. Further, using the Cauchy-Schwartz inequality and the sub-multiplicative property of induced matrix norms,
\begin{align*}
\xi_{k}^{T}\mathcal{A}^TP\eta_k &= \left(\mathcal{A}\xi_k\right)^TP\eta_k \\
&\leq \left\|\mathcal{A}\xi_k\right\|\left\|P\eta_k\right\| \\
&\leq \left\|\xi_k\right\|\left\|\mathcal{A}\right\|\left\|P\right\|\left\|\eta_k\right\| \\
&= \left\|\xi_k\right\|\left\|\mathcal{A}\right\|p_{\mathrm{max}}\left\|\eta_k\right\|,
\end{align*}
as $P$ is symmetric and positive definite. Using the above results in \eqref{eq:Lyapunov1},
\begin{equation} \label{eq:Lyapunov2}
\Delta V_k \leq  -\left\|\xi_k\right\|^2 + 2\left\|\xi_k\right\|\left\|\mathcal{A}\right\|p_{\mathrm{max}}\left\|\eta_k\right\| + p_{\mathrm{max}}\left\|\eta_k\right\|^2.
\end{equation}
For asymptotic stability, it is required that $\Delta V_k < 0$. This is achieved when
\begin{equation} \label{eq:Lyapunov3}
\left\|\xi_k\right\|^2 - 2\left\|\mathcal{A}\right\|p_{\mathrm{max}}\left\|\eta_k\right\|\left\|\xi_k\right\| - p_{\mathrm{max}}\left\|\eta_k\right\|^2 > 0.
\end{equation}
This is a simple quadratic equation in $\left\|\xi_k\right\|$. Solving, $\Delta V_k < 0$ if:
\begin{equation} \label{eq:QCond}
\left\|\xi_k\right\| > \left\|\eta_k\right\|\left[\left\|\mathcal{A}\right\|p_{\mathrm{max}} + \sqrt{\left\|\mathcal{A}\right\|^2p_{\mathrm{max}}^2 + p_{\mathrm{max}}}\right],
\end{equation}
i.e. $\xi_k$ converges exponentially to the bounded ball of radius $\mathcal{R}$, where
\begin{equation} \label{eq:RadConv}
\mathcal{R} = \left\|\eta_k\right\|\left[\left\|\mathcal{A}\right\|p_{\mathrm{max}} + \sqrt{\left\|\mathcal{A}\right\|^2p_{\mathrm{max}}^2 + p_{\mathrm{max}}}\right].
\end{equation}
Note that with slowly varying $L_d(k)$, $\left\|\eta_k\right\|$ is small, and $\xi_k$ converges to a bounded ball of smaller radius, closer to the origin. This quantitatively validates the earlier statement on asymptotic stability with slowly varying $L_d(k)$.

\begin{figure*}[t!]
\begin{center}
\begin{tikzpicture}[auto,>=latex',connect with angle/.style=
{to path={let \p1=(\tikztostart), 
\p2=(\tikztotarget) in -- ++({\x2-\x1-(\y1-\y2)*tan(#1-90)},0) -- (\tikztotarget)}}]

    \node [block, align = center] (plant) {Plant with Uncertainties\\$\dot{x}(t) = Ax(t) + Bu(t) + L(t)$};
    \node [block, right of = plant, node distance = 4.5cm] (sampler1) {Sampler};
    \node [square, left of = plant, node distance = 3.8cm] (ZOH) {ZOH};
    \node [sum, below right = 0.15cm and 0.85cm of sampler1] (Sum1) {\Large$\Sigma$};
    \node [block, below of = plant, node distance = 2cm, align = center] (observer) {Observer\\$\hat{x}(k+1) = F_n\hat{x}(k) + G_nu(k) + \hat{D}_d(x, k) + L(y(k) - \hat{y}(k))$};
    \node [coord, right of = observer, node distance = 6.72cm] (interobs) {};
    \node [coord, left of = observer, node distance = 5.461cm] (obsinput) {};
    \node [sum, below of = Sum1, node distance = 2cm] (Sum2) {\Large$\Sigma$};
    \node [coord, below of = Sum2, node distance = 1.01cm] (interref1) {};
    \node [coord, right of = Sum1, node distance = 2cm] (eSE) {};
    \node [coord, right of = Sum2, node distance = 1.5cm] (inter1) {};
    \node [block, below of = observer, node distance = 2cm, align = center] (refmod) {Reference Model\\$\dot{x}_m(t) = A_mx_m(t) + B_mr(t)$};
    \node [block, right of = refmod, node distance = 4.5cm] (sampler2) {Sampler};
    \node [block, below right = 0.5cm and -3cm of refmod] (controller) {Controller};
    \node [coord, left of = refmod, node distance = 3.75cm] (ref) {};
    \node [coord, left of = controller, node distance = 5.8cm] (inter2) {};

    \draw[-latex] (ref) -- node{$r(k)$} (refmod);
    \draw[-latex] (plant) -- node{$x(t)$} (sampler1);
    \draw[-latex] (sampler1) -| node[xshift = -0.9cm, yshift = 0.25cm]{$x(k)$} node[yshift = -0.4cm] {$+$} (Sum1);
    \draw[-latex] (refmod) -- node{$x_m(t)$} (sampler2);
    \draw[-latex] (sampler2) -| node[yshift = 0.25cm]{$x_m(k)$} node[xshift = 0.55cm, yshift = 0.45cm] {$-$} (Sum2);
    \draw[-] (observer) -- node[yshift = -0.05cm]{$\hat{x}(k)$} (interobs);
    \draw[-latex] (interobs) -- node[xshift = 0.55cm, yshift = 0.15cm]{$-$} (Sum1);
    \draw[-latex] (interobs) -- node[yshift = -0.1cm]{$+$} (Sum2);
    \filldraw[black] (interobs) circle (1.5pt) node{};
    \draw[-latex] (Sum1) -- node{$e_{SE}(k)$} (eSE);
    \draw[-] (Sum2) -- node{$\hat{e}(k)$} (inter1);
    \draw[-latex, name path = line1] (inter1) |- node[xshift = -6.25cm, yshift = 0.6cm]{$\hat{e}(k)$} (controller);
    \draw[-] (controller) -- node[xshift = 1.9cm, yshift = 0.6cm]{$u(k)$} (inter2);
    \draw[-latex] (inter2) |- node[xshift = 0.6cm, yshift = 0cm]{$u(k)$} (ZOH);
    \draw[-latex] (ZOH) -- node{$u(t)$} (plant);
    \draw[-latex] (obsinput) -- node{$u(k)$} (observer);
    \filldraw[black] (obsinput) circle (1.5pt) node{};

\end{tikzpicture}
\end{center}
\caption{Block diagram for controlling a continuous-time system using discrete-time UDE-based controller--observer structure.}
\label{fig:CT-DTObs}
\hrulefill
\end{figure*}
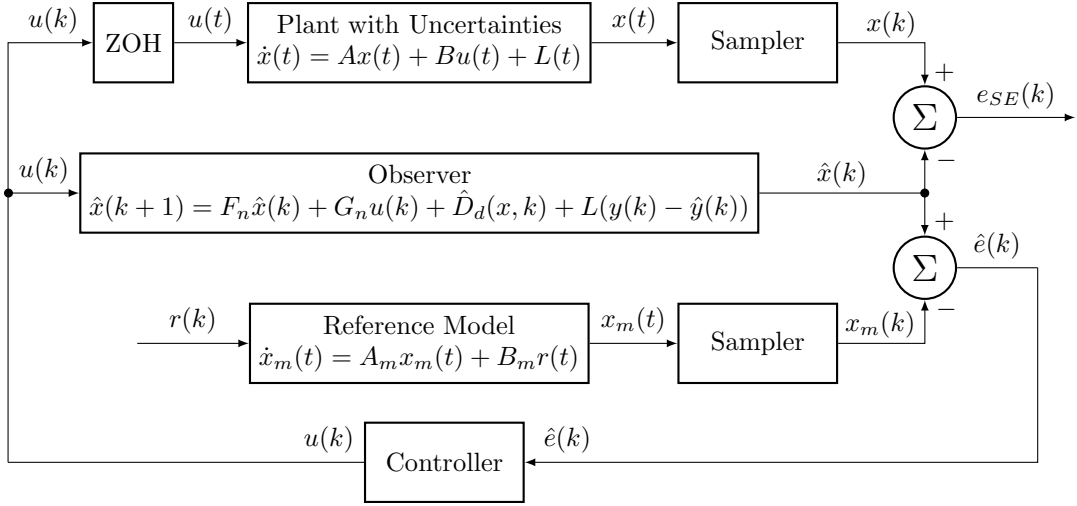

Summarizing the results in this section, a robust discrete-time observer has been designed by modifying a simple Luenberger observer by including the estimate of the disturbance affecting the plant \eqref{eq:PlantObs}. As the control law in \eqref{eq:Control} depends on the unavailable state tracking error $e(k)$, an auxiliary error $\hat{e}(k) = \hat{x}(k) - x_m(k)$, generated using the state estimate, is used to generate a new control law in \eqref{eq:Control-Observer}. This results in a DT-UDE-based controller--observer structure. A simple, qualititative analysis of stability demonstrates that given $\big\{C, F_n, G_n\big\}$ is a minimal realization, the controller and observer gains $K_d$ and $\beta$, along with the filter parameter $\tau$ can be chosen to ensure stability of the controller--observer structure. Further, a quantitative analysis of stability provides bounds on the energy of the overall error vector $\xi(k)$, consisting of the auxiliary error $\hat{e}(k)$, the state estimation error $e_{SE}(k)$ and the disturbance estimation error $\Tilde{L}_d(k)$.

As before, the entire analysis above has been for a discrete-time plant \eqref{eq:PlantObs}. To control a continuous-time plant such as the robot manipulator \eqref{eq:RMDynSSMat}, the continuous-time reference model states $x_m(t)$ must be sampled to obtain $x_m(k)$ and subsequently the auxiliary error $\hat{e}(k)$. This is used in the controller to generate $u(k)$ to drive the original continuous-time plant using a zero-order-hold. This procedure is illustrated in Fig.~\ref{fig:CT-DTObs}. Note how the plant is a continuous-time system, but the observer remains a discrete-time system. For simplicity, the tracking error $e(k) = x(k) - x_m(k)$ is not shown, but it is simply given by $e(k) = e_{SE}(k) + \hat{e}(k)$. The differences between Figures \ref{fig:DT-UDE-Observer} and \ref{fig:CT-DTObs} are evident. The reference model state vector $x_m(t)$ is sampled, resulting in $x_m(k)$, which is used along with the observer state vector $\hat{x}(k)$ to generate the auxiliary error $\hat{e}(k)$ and in turn the control input $u(k)$ in \eqref{eq:Control-Observer}. While $u(k)$ drives the discrete-time observer directly, the continuous-time plant is driven by passing $u(k)$ through a zero-order-hold. 

Finally, an interesting avenue would be to explore modifications to $G(\gamma)$ to mitigate the effect of $\Delta L_d(k)$ in \eqref{eq:ObsDyn}. This is particularly important in the general case of an $n$-link manipulator, where the condition $\Delta L_d(k) \approx 0$ may not hold. Progress has been made towards this in the continuous-time case, in particular the use of second-order or higher-order filters for disturbance estimation \cite{IOL-UDE}, and modification of the first-order filter using a new parameter $\alpha$ \cite{Alpha}. Both avenues have shown promising performance, and exploring the equivalent techniques in discrete-time is an interesting avenue for future work.

\begin{figure}
\centering

\begin{subfigure}{0.42\textwidth}
	\includegraphics[width = \textwidth]{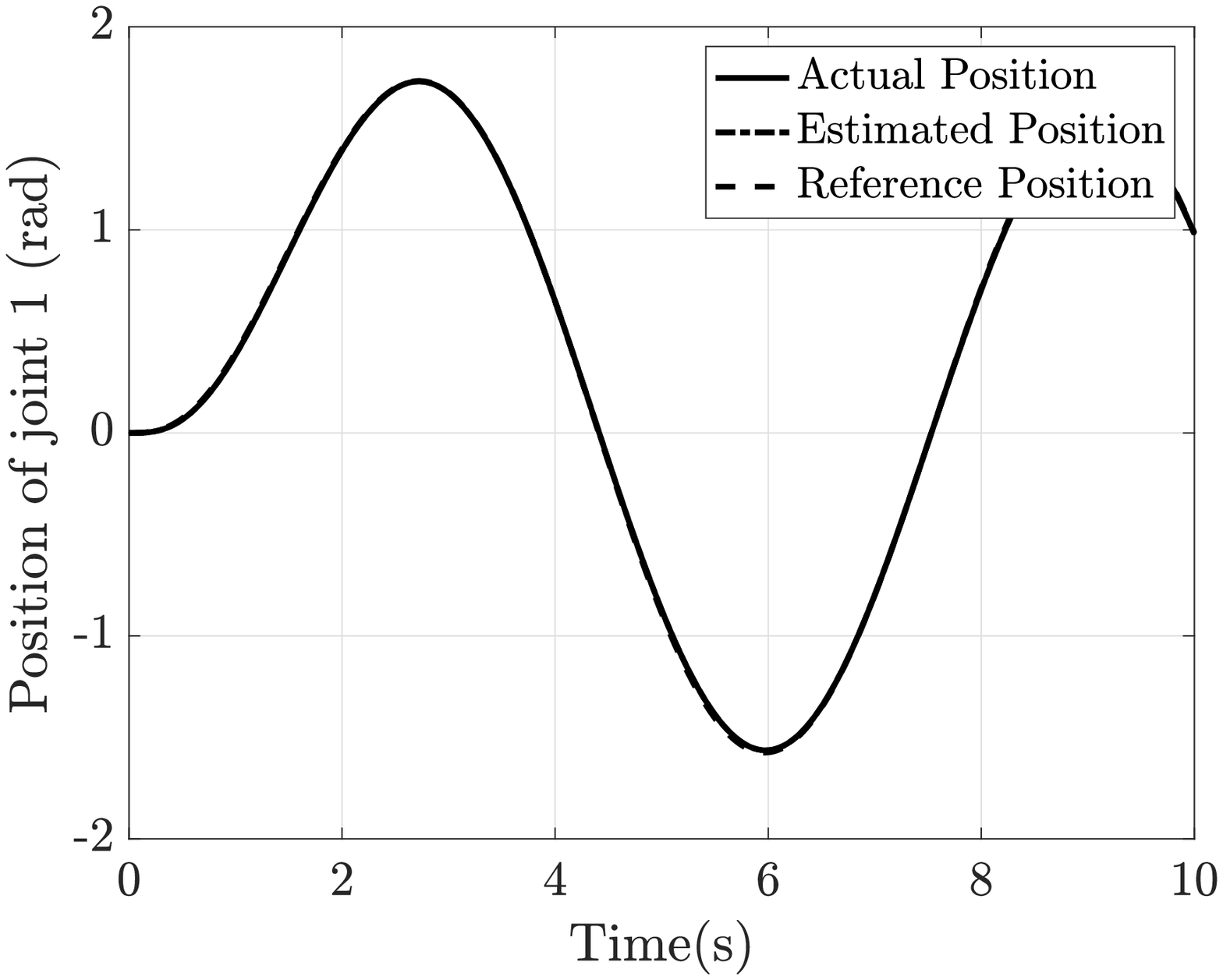}
	\caption{Position of link $1$}
\end{subfigure}
\hfill
\begin{subfigure}{0.42\textwidth}
	\includegraphics[width = \textwidth]{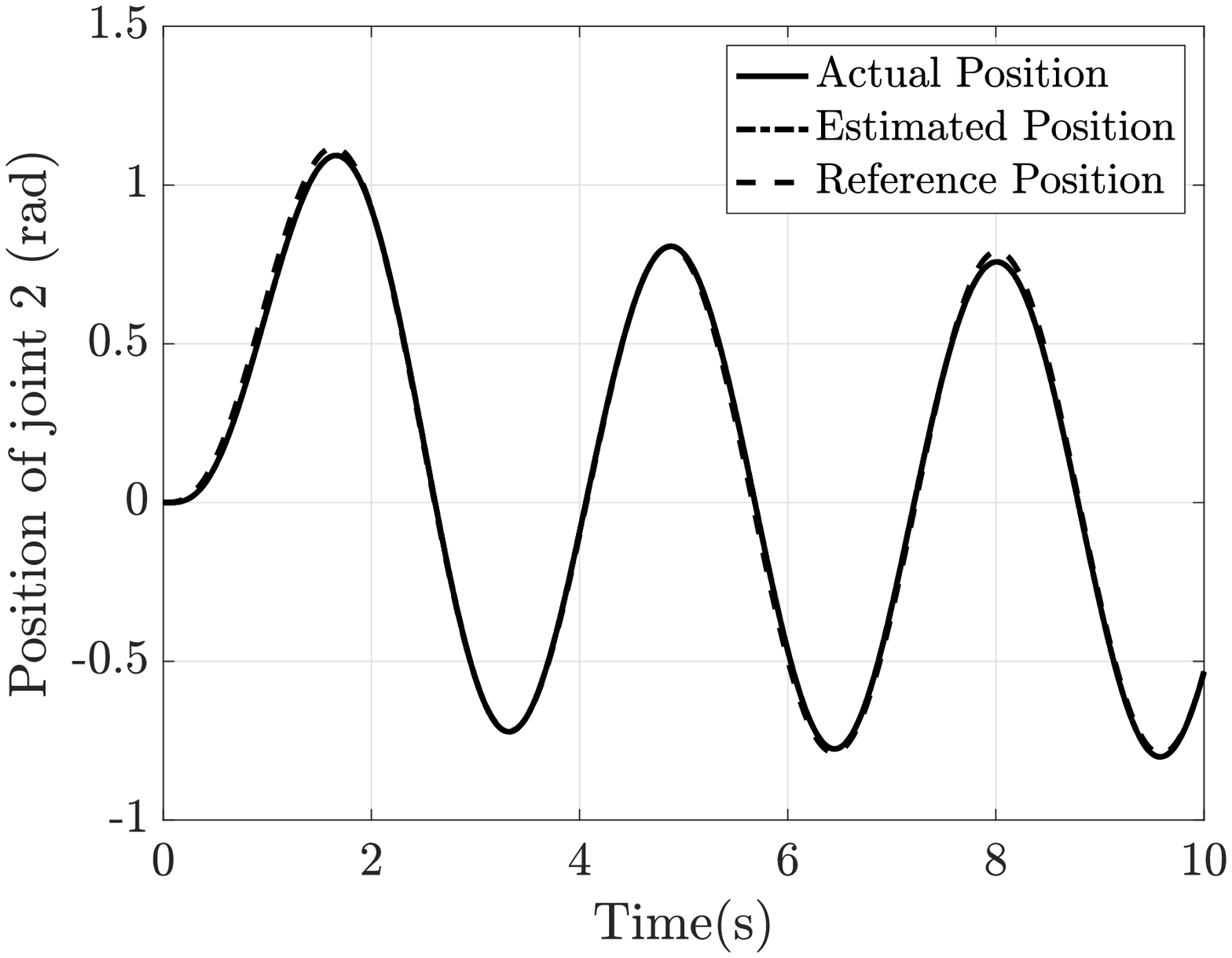}
	\caption{Position of link $2$}
\end{subfigure}
\\
\begin{subfigure}{0.42\textwidth}
	\includegraphics[width = \textwidth]{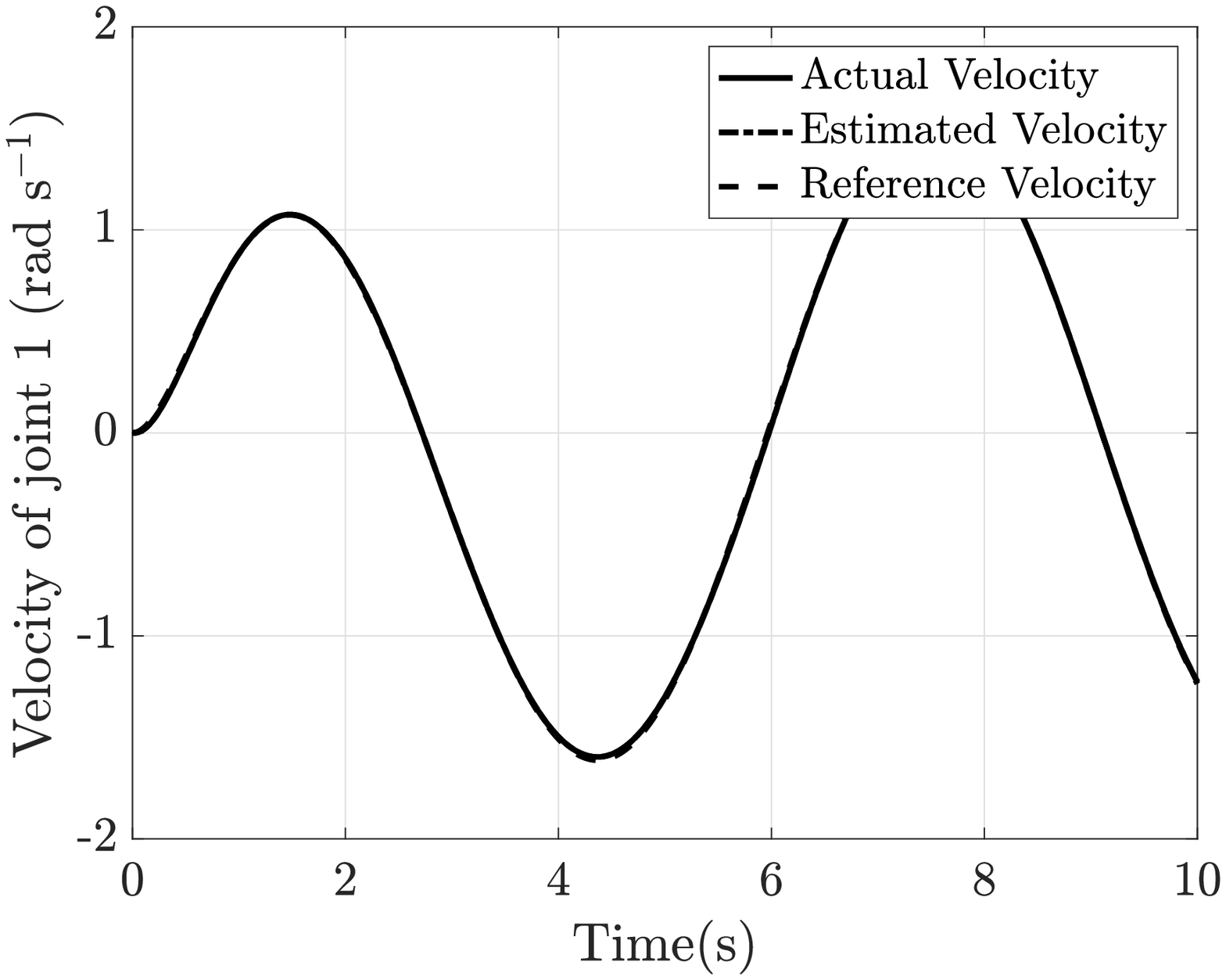}
	\caption{Velocity of link $1$}
\end{subfigure}
\hfill
\begin{subfigure}{0.42\textwidth}
	\includegraphics[width = \textwidth]{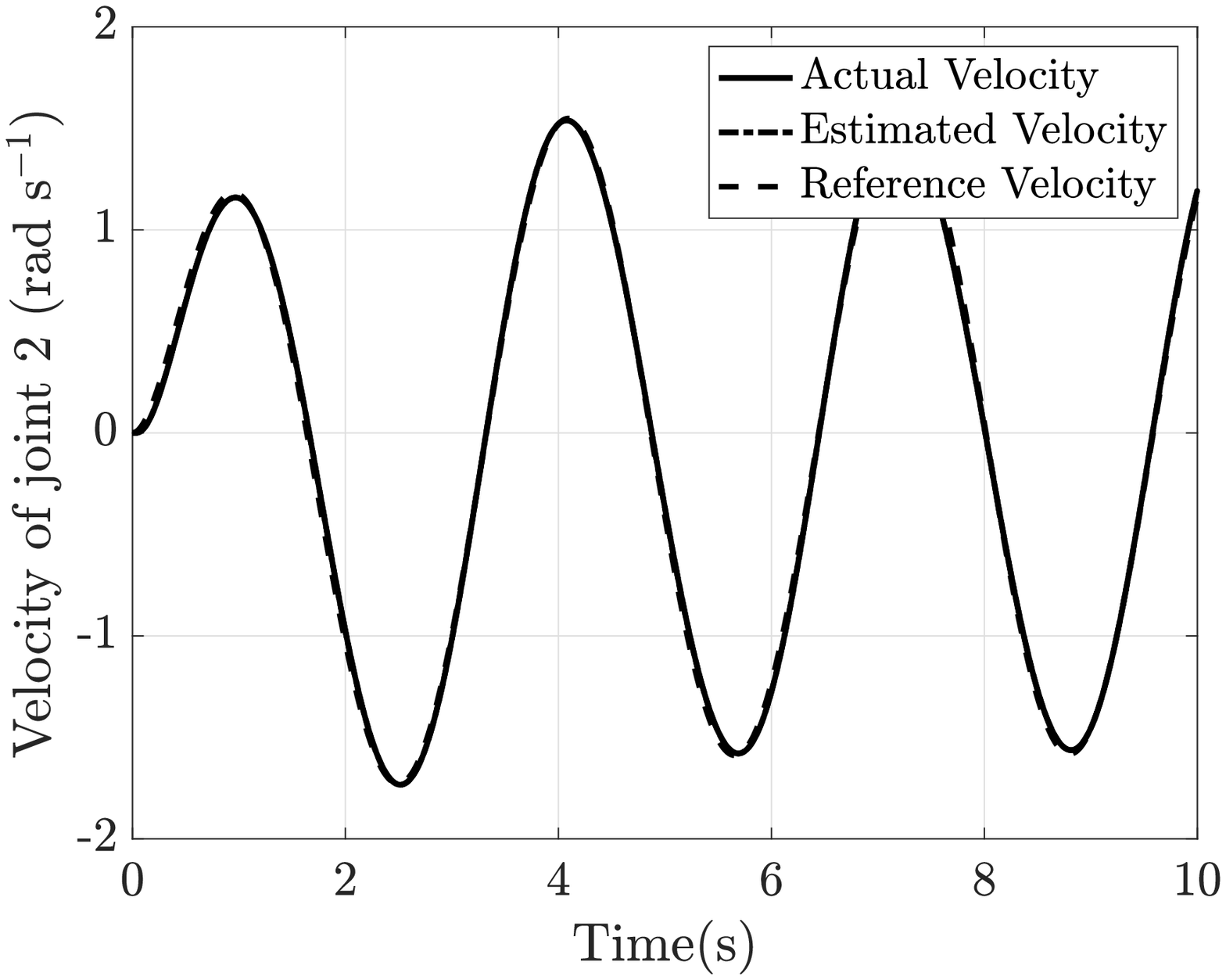}
	\caption{Velocity of link $2$}
\end{subfigure}
\\
\begin{subfigure}{0.42\textwidth}
	\includegraphics[width = \textwidth]{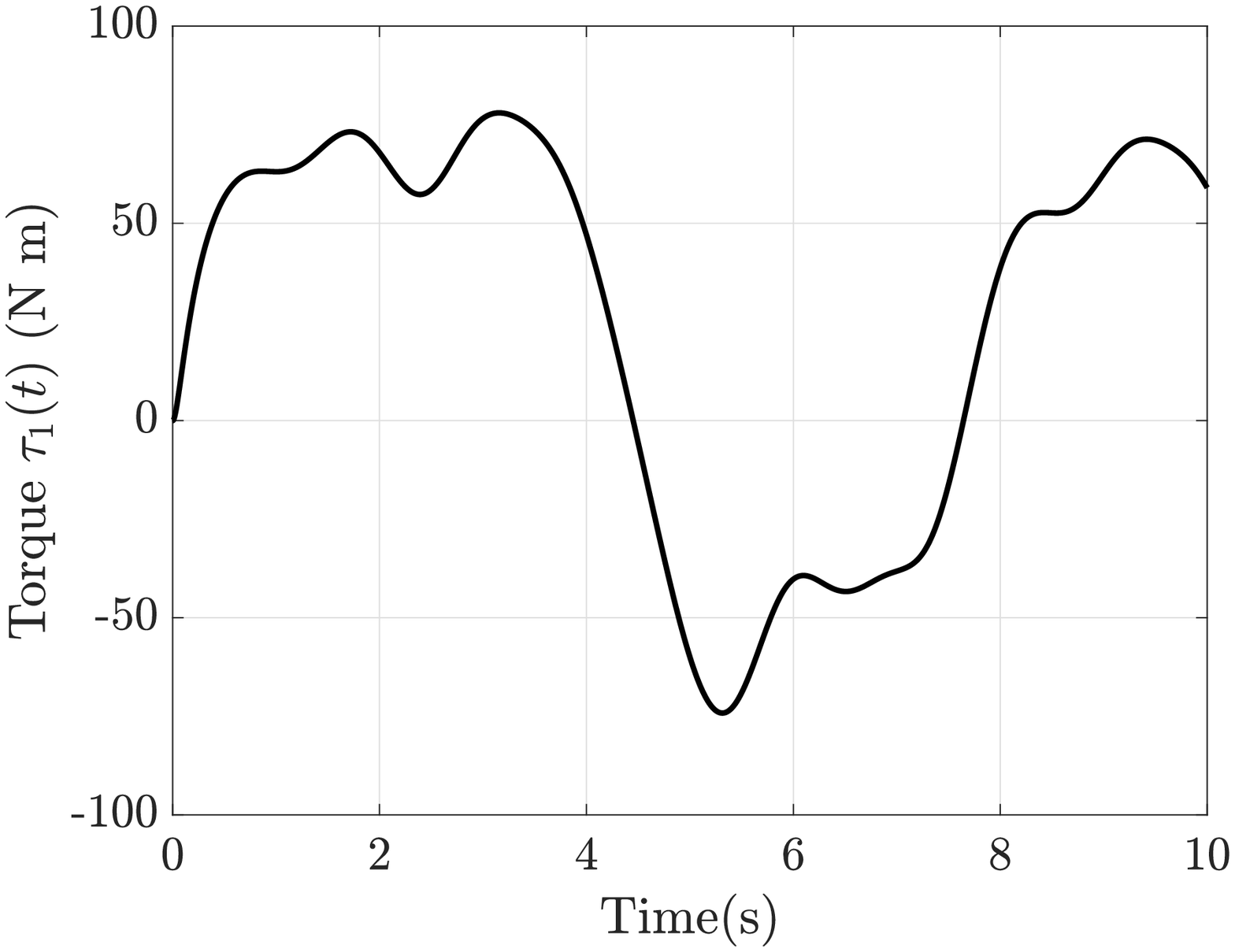}
	\caption{Torque on link $1$}
\end{subfigure}
\hfill
\begin{subfigure}{0.42\textwidth}
	\includegraphics[width = \textwidth]{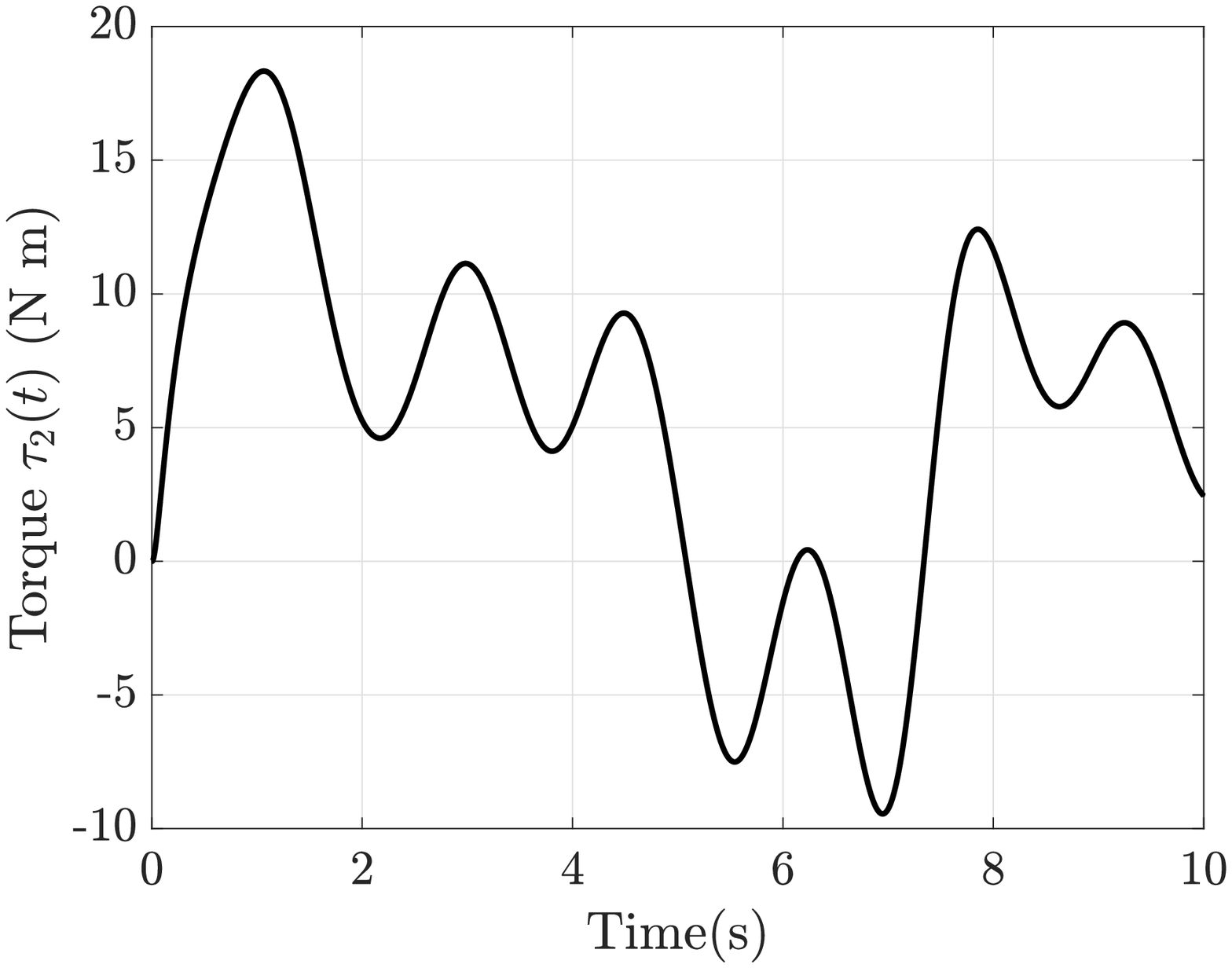}
	\caption{Torque on link $2$}
\end{subfigure}
\\
\begin{subfigure}{0.42\textwidth}
	\includegraphics[width = \textwidth]{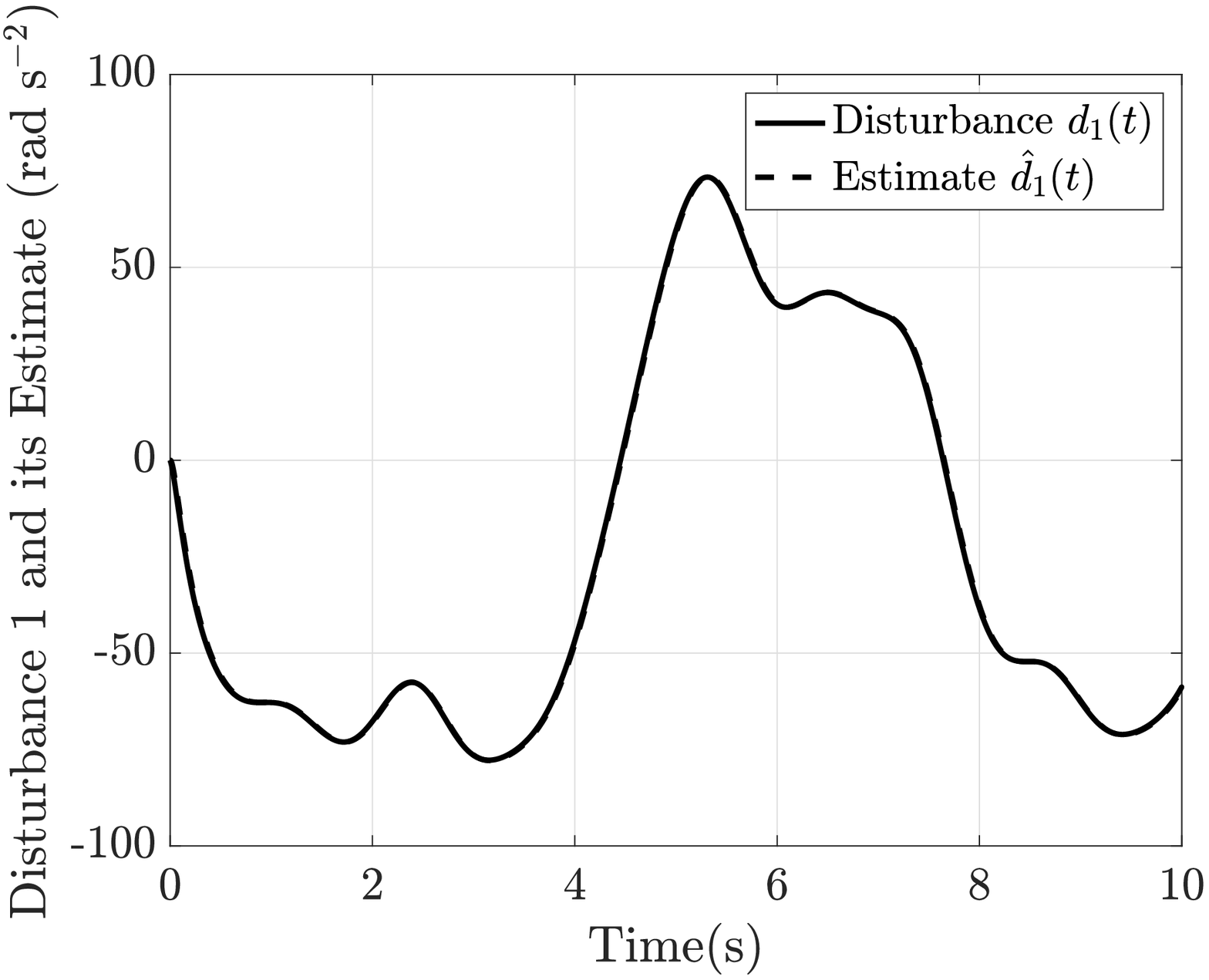}
	\caption{Disturbance $d_1$ and its estimate}
\end{subfigure}
\hfill
\begin{subfigure}{0.42\textwidth}
	\includegraphics[width = \textwidth]{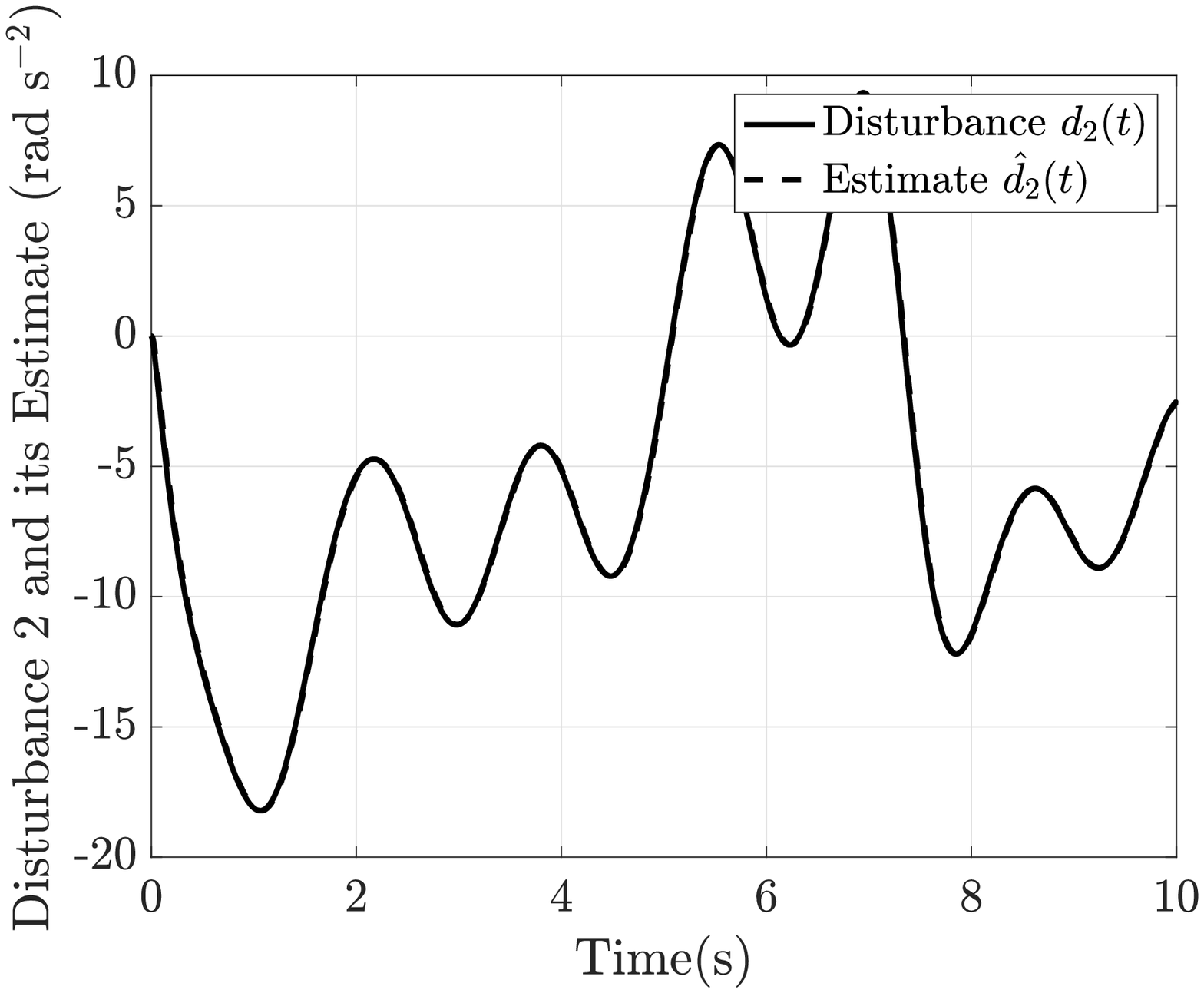}
	\caption{Disturbance $d_2$ and its estimate}
\end{subfigure}

\caption{Simulation results for controlling the robot manipulator system using the robust discrete-time observer in Section \ref{sec:Observer}.}
\label{fig:Sim-DT-UDE}
\hrulefill
\end{figure}

\begin{table}[!t]
\centering
\caption{Simulation Parameters}
\begin{tabular}{lll} \toprule
Parameter & Symbol & Value \\ \midrule
Mass of link 1 & $m_1$ & $2$ kg \\
Mass of link 2 & $m_2$ & $1$ kg \\
Length of link 1 & $l_1$ & $2$ m \\
Length of link 2 & $l_2$ & $1$ m \\ 
Uncertain mass $m_1$ & $m_{1u}$ & $2.4$ kg \\
Uncertain mass $m_2$ & $m_{2u}$ & $1.3$ kg \\
Uncertain length $l_1$ & $l_{1u}$ & $2.5$ m \\
Uncertain length $l_2$ & $l_{2u}$ & $1.2$ m \\
Acceleration due to gravity & $g$ & 9.8 m s$^{-2}$ \\ 
Filter parameter & $\tau$ & 0.01 \\
\multirow{2}{*}{Reference Model Inputs} & $r_1(t)$ & $5\sin(t)$ \\
& $r_2(t)$ & $5\sin(2t)$ \\ \bottomrule
\end{tabular}
\label{tab:Parameters}
\end{table}

\section{Simulation Results} \label{sec:Simulations}

\subsection{Performance of robust controller--observer structure} \label{sec:Sim1}
In this section, simulation results for the robust discrete-time UDE-based controller--observer structure presented in Section \ref{sec:Observer}, applied to the robot manipulator system in \eqref{eq:RMDynSSMat} are presented. The actual and uncertain system parameters are available in Table \ref{tab:Parameters}. It can be seen that the uncertainty in parameters is chosen to be between $20\%$ and $30\%$ for the simulations. Further, an external state-dependent disturbance $d' = \begin{bmatrix} 20\sin 2\pi\theta_1(t) \\ 10\sin 2\pi\theta_2(t) \end{bmatrix}$ is chosen to affect the system. The objective is to ensure that the states of the robot manipulator system in \eqref{eq:RMDynSSMat} tracks the states of the reference model in \eqref{eq:CTRefModel}, with matrices $A_m$ and $B_m$ given by:
$$
A_m = \begin{bmatrix} \phantom{-}0 & \phantom{-}1 & \phantom{-}0 & \phantom{-}0 \\
-2 & -3 & \phantom{-}0 & \phantom{-}0 \\
\phantom{-}0 & \phantom{-}0 & \phantom{-}0 & \phantom{-}1 \\
\phantom{-}0 & \phantom{-}0 & -2 & -3 \end{bmatrix}; \hspace{0.2cm} 
B_m = \begin{bmatrix} 0 & 0 \\ 1 & 0 \\ 0 & 0 \\ 0 & 1 \end{bmatrix}
$$
and the inputs to the reference model $r_1(t)$ and $r_2(t)$ given in Table \ref{tab:Parameters}. The state tracking error dynamics as well as the dynamics of the reference model are discretized with a sampling time $T_s = 0.01\mathrm{ s}$, to apply the robust controller--observer structure. This results in matrices $F_n$, $G_n$, $F_m$ and $G_m$ for the plant and reference model, with expressions given in Section \ref{sec:UDEReview}. The feedback gain $K_d$ for the control law is designed such that the eigenvalues of $\left(F_n - G_nK_d\right)$ coincide with the eigenvalues of $F_m$. The observer gain matrix $\beta$ is chosen such that the eigenvalues of $\left(F_n - \beta C\right)$ are chosen at $0.1$ times the eigenvalues of $F_m$, i.e. much closer to the origin. This results in $K_d = \begin{bmatrix} 33.95 & 50.95 & 0 & 0 \\ 0 & 0 & 1.99 & 2.99 \end{bmatrix}$ and $\beta = \begin{bmatrix} 1.8 & 810.27 & 0 & 0 \\ 0 & 0 & 1.8 & 810.27 \end{bmatrix}^T$. The initial conditions on the plant, reference model and observer are all chosen to be zero. The control law $u(k)$ in \eqref{eq:Control-Observer} is designed, and passed through a zero-order-hold before driving the original robot manipulator system. Using the parameters listed in Table \ref{tab:Parameters}, the results in Fig. \ref{fig:Sim-DT-UDE} are obtained. From Fig. \ref{fig:Sim-DT-UDE}(a)-(d), it is evident that the robust observer designed accurately estimates system states, leading to accurate tracking performance. The torque inputs remain at acceptable levels in Fig. \ref{fig:Sim-DT-UDE}(e)-(f), and the digital filter $G(\gamma)$ models the `lumped' disturbance with high accuracy in Fig. \ref{fig:Sim-DT-UDE}(g)-(h). In summary, the DT-UDE-based controller--observer structure from Section \ref{sec:Observer} achieves the required tracking objective with high accuracy.

\subsection{Comparison with Existing Designs} \label{sec:Sim2}
In this section, an extensive comparative study of different strategies for controlling robot manipulators is presented. Three well-known strategies are considered: (1) Sliding-Mode Control (SMC), (2) Gravity-compensated proportional-derivative (PD) control, and (3) Continuous-time UDE-based control. These three designs are compared with the proposed DT-UDE-based controller--observer structure. Throughout this study, desired output settling times of 0.5 seconds, with critically damped/overdamped characteristics are chosen. The three comparative designs are briefly outlined below.

\subsubsection{Design 1: Sliding-Mode Control (SMC)}
The first strategy considered is Sliding-Mode Control (SMC). As described in Section \ref{sec:Introduction}, SMC has a rich history in the robust control of robot manipulators, and the design considered here is based on the description in \cite{Survey1}. The control law is given by:
\begin{equation} \label{eq:SMC1}
	\tau = -Ke + \tau_{ff} + \tau_d.
\end{equation}
Here, $e(t) = x(t) - x_m(t)$ denotes the state tracking error, $\tau_{ff}(t)$ denotes the feedforward component of $\tau$, and $\tau_d(t)$ denotes the robust control component of $\tau$. $K$ is chosen according to the usual principles of state feedback, such that the eigenvalues of $(A - BK)$ coincide with the eigenvalues of $A_m$. This results in $K = \begin{bmatrix} 34 & 51 & 0 & 0 \\ 0 & 0 & 2 & 3 \end{bmatrix}$. The feedforward component is given by:
\begin{equation} \label{eq:Tauff}
	\tau_{ff} = \hat{M}(\theta)\Ddot{\theta}_m + \hat{C}(\theta, \dot{\theta}) + \hat{K}(\theta),
\end{equation}
where $\hat{M}(\theta)$, $\hat{C}(\theta, \dot{\theta})$ and $\hat{K}(\theta)$ denote the inertia matrix, Coriolis component and gravitational component constructed based on the uncertain parameters in Table \ref{tab:Parameters}. $\Ddot{\theta}_m$ is the second derivative of reference joint positions, i.e. $\displaystyle \Ddot{\theta}_m = \begin{bmatrix} \Ddot{\theta}_{1m} \\ \Ddot{\theta}_{2m} \end{bmatrix}$. The robust control component is given by:
\begin{equation} \label{eq:Taud}
	\tau_d = D.\mathrm{sat}\left(s/\epsilon\right).
\end{equation}
$D$ represents the assumed bound on uncertainty, $s$ denotes the sliding surface:
\begin{equation} \label{eq:Sliding}
	s = \dot{e} + K_De,
\end{equation}
$\mathrm{sat}(.)$ denotes the saturation function:
\begin{equation} \label{eq:Saturation}
	\mathrm{sat}\left(s/\epsilon\right) = 
	\begin{dcases}
	\frac{s}{\|s\|} & \text{if } \|s\| > \epsilon \\
	\frac{s}{\epsilon} & \text{else}
	\end{dcases},
\end{equation}
and $\epsilon$ denotes the width of the boundary layer. Throughout the simulations, the bounds on uncertainties $d_1$ and $d_2$ are taken to be $15$ and $10$, $\epsilon$ is chosen to be $0.1$ and $K_D = 7$.

\subsubsection{Design 2: Gravity-compensated PD-control}
In this design, a proportional-derivative (PD) controller compensated by the gravitational force matrix $K(\theta)$ is designed as follows \cite{Spong}:
\begin{equation} \label{eq:GPD}
	\tau = K_P\left(\theta - \theta_m\right) + K_D\left(\dot{\theta} - \dot{\theta}_m\right) + K(\theta),
\end{equation}
where $\theta$ denotes the output vector, i.e. the vector of link positions, $\dot{\theta}$ denotes the vector of link velocities, and $\theta_m$ and $\dot{\theta}_m$ denote the desired trajectories for the link positions and velocities respectively, obtained from the reference model. For the set-point control of robot manipulators, it has been shown that such a controller offers robust tracking performance \cite{Survey1}. For the simulations, $K_P = 1$ and $K_D = 0.1$ are chosen to satisfy the desired performance specifications.

\subsubsection{Design 3: Continuous-time UDE-based control}
The third design strategy considered is the continuous-time UDE-based control, first proposed in \cite{CT-UDE}. This was also applied to the control of robot manipulators in \cite{CT-UDE-RM}. The control law applied is:
\begin{equation}
	u(t) = -Ke(t) -\frac{1}{\tau}B^{\dagger}\left[e(t) - \big(A - BK\big)\int e(t)\mathrm{d}t\right],
\end{equation}
as derived in \cite{DT-UDE}. $K$ is chosen to ensure the eigenvalues of $(A-BK)$ coincide with the eigenvalues of the reference model $A_m$, resulting in $K = \begin{bmatrix} 34 & 51 & 0 & 0 \\ 0 & 0 & 2 & 3 \end{bmatrix}$.

\subsubsection{Design 4: Discrete-time UDE-based control}
The final design strategy is the proposed strategy in this article, the DT-UDE-based controller--observer structure. In contrast to the approach in Section \ref{sec:Sim1}, $K_d$ is chosen to satisfy the desired performance specifications, and the observer gain matrix is chosen such that the eigenvalues of $\left(F_n - \beta C\right)$ are at $0.1$ times the eigenvalues of $\left(F_n - G_nK_d\right)$. This results in the same values of $K_d$ and $\beta$ as mentioned in Section \ref{sec:Sim1}. As before, $u(k)$ in \eqref{eq:Control-Observer} is designed and passed through a zero-order-hold before driving the original robot manipulator system.

\begin{figure*}[!t]
\centering

\begin{subfigure}{0.4\textwidth}
	\centering
	\includegraphics[width = \textwidth]{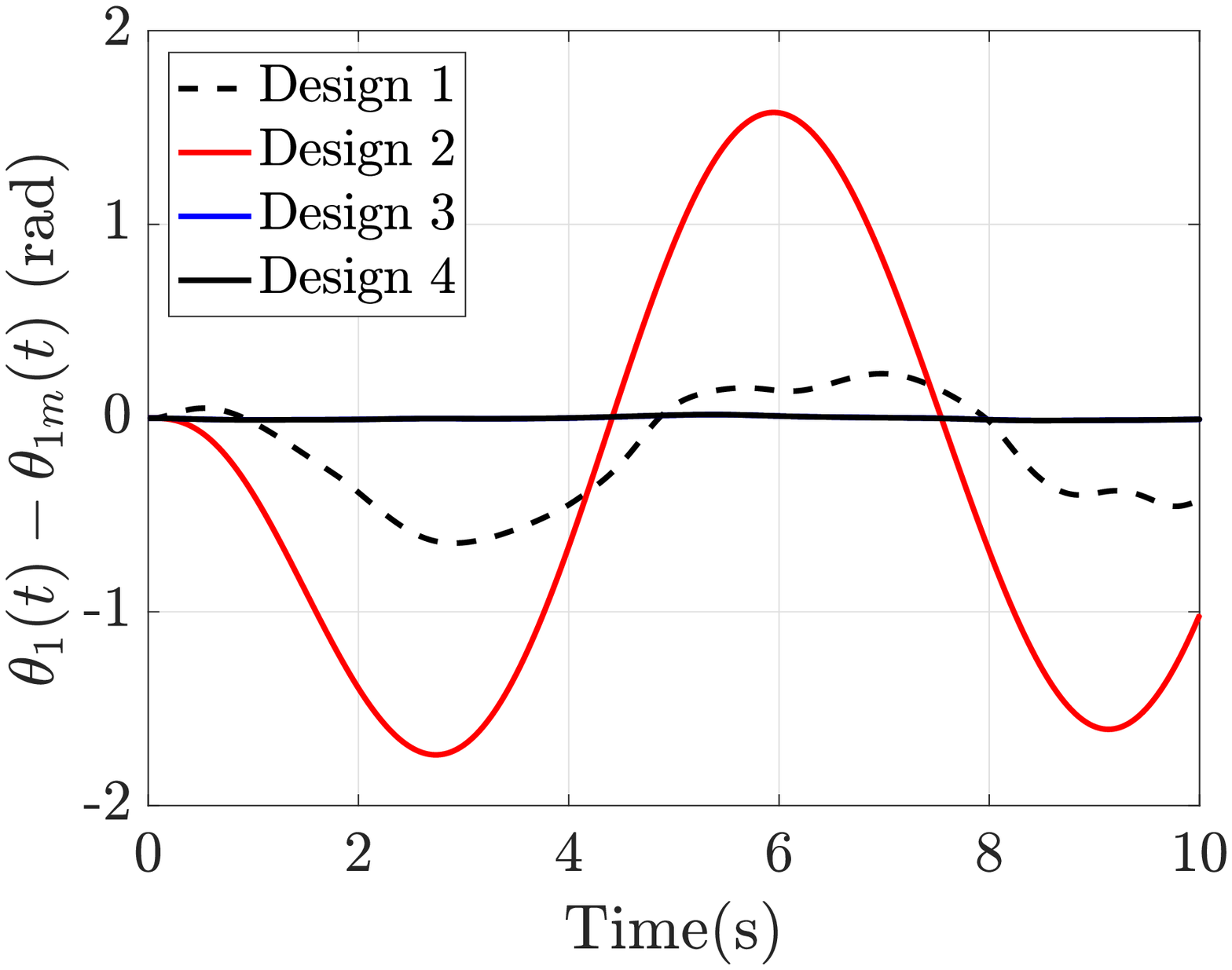}
	\caption{Position tracking error in link $1$}
\end{subfigure} \hfill
\begin{subfigure}{0.4\textwidth}
	\centering
	\includegraphics[width = \textwidth]{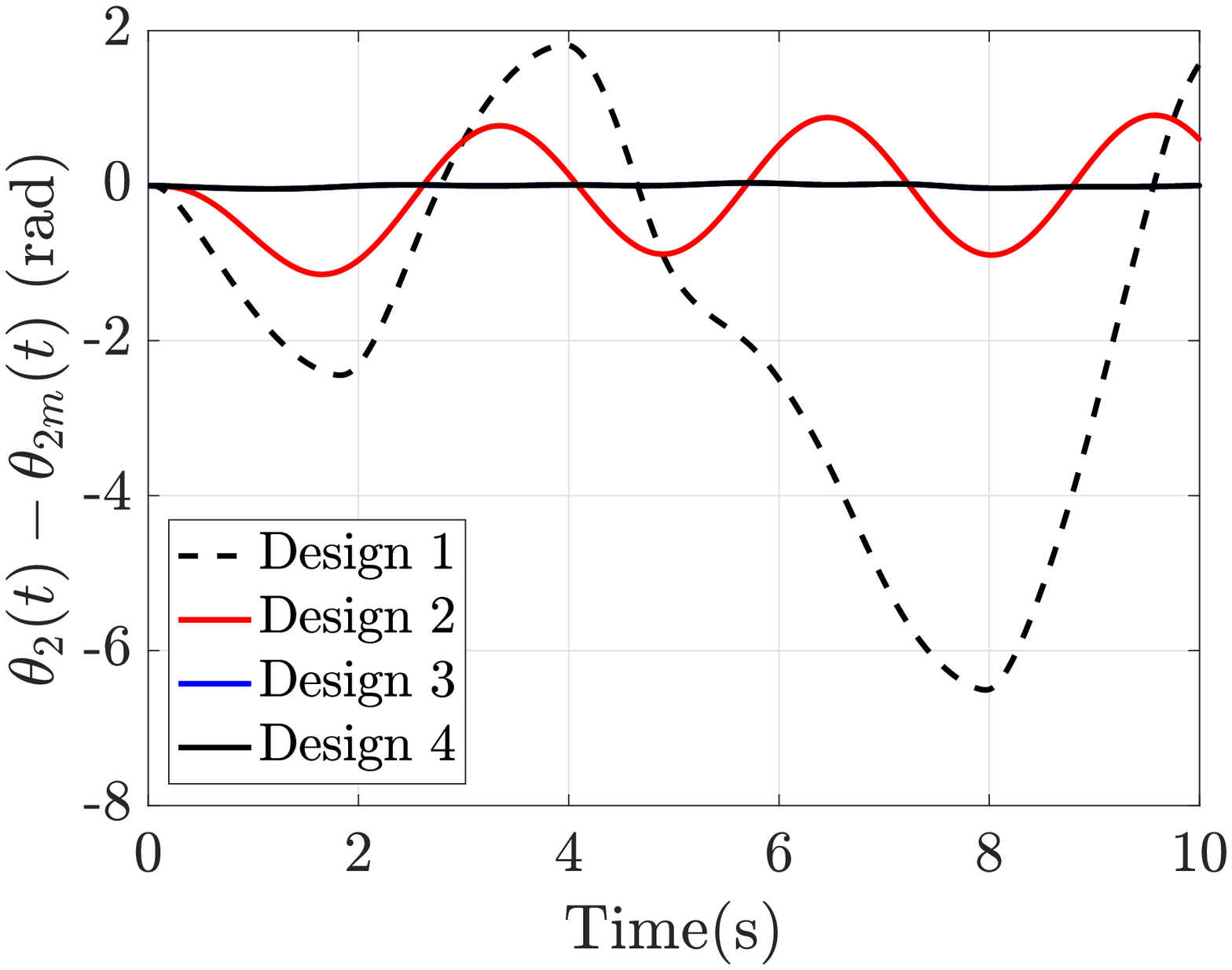}
	\caption{Position tracking error in link $2$}
\end{subfigure}

\caption{Output tracking error comparisons across the four designs.}
\label{fig:Comparison}
\hrulefill

\end{figure*}

\begin{figure*}[!t]
\centering

\begin{subfigure}{0.4\textwidth}
	\centering
	\includegraphics[width = \textwidth]{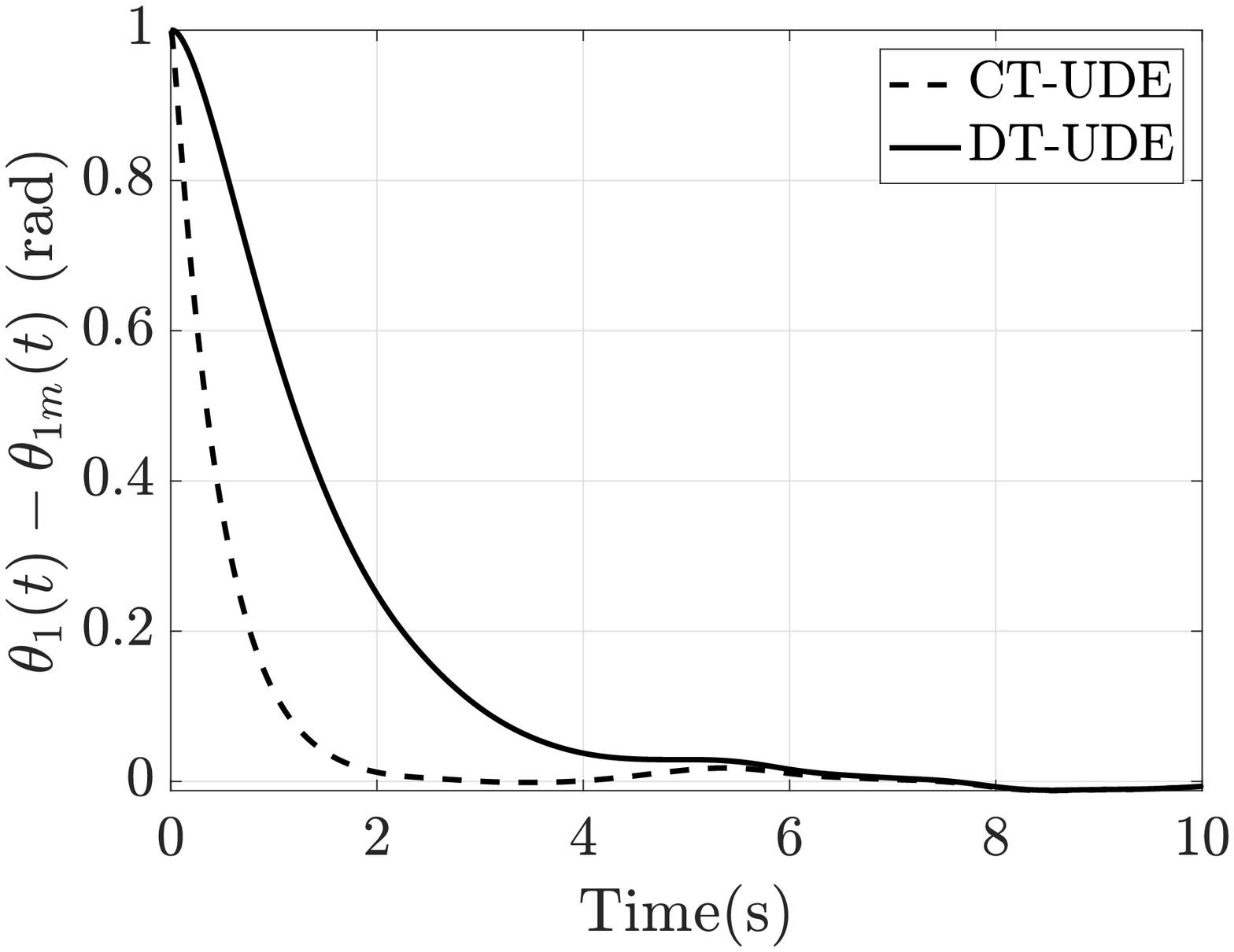}
	\caption{Position tracking error in link $1$}
\end{subfigure} \hspace{2cm}
\begin{subfigure}{0.4\textwidth}
	\centering
	\includegraphics[width = \textwidth]{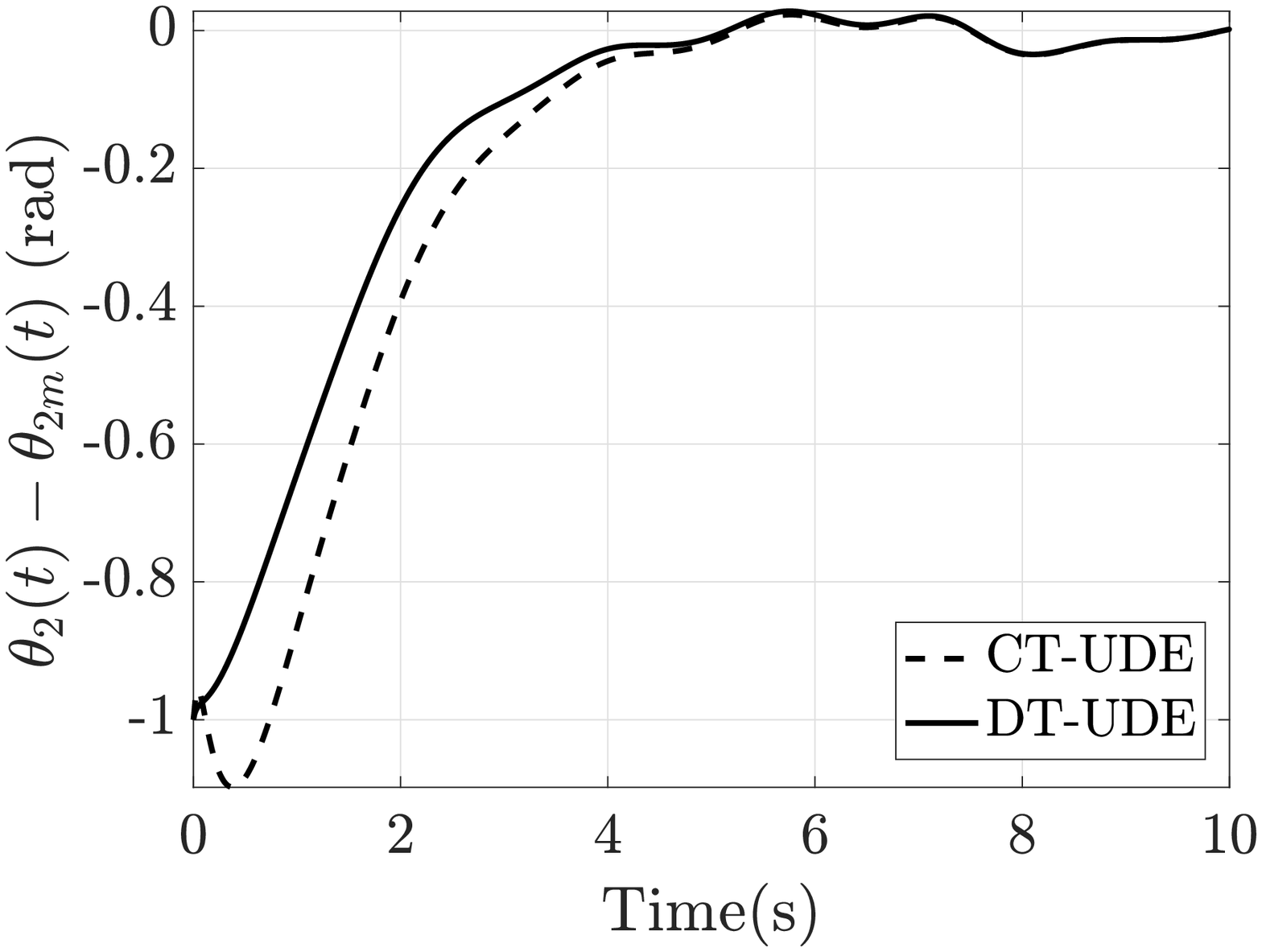}
	\caption{Position tracking error in link $2$}
\end{subfigure} \\
\begin{subfigure}{0.4\textwidth}
	\centering
	\includegraphics[width = \textwidth]{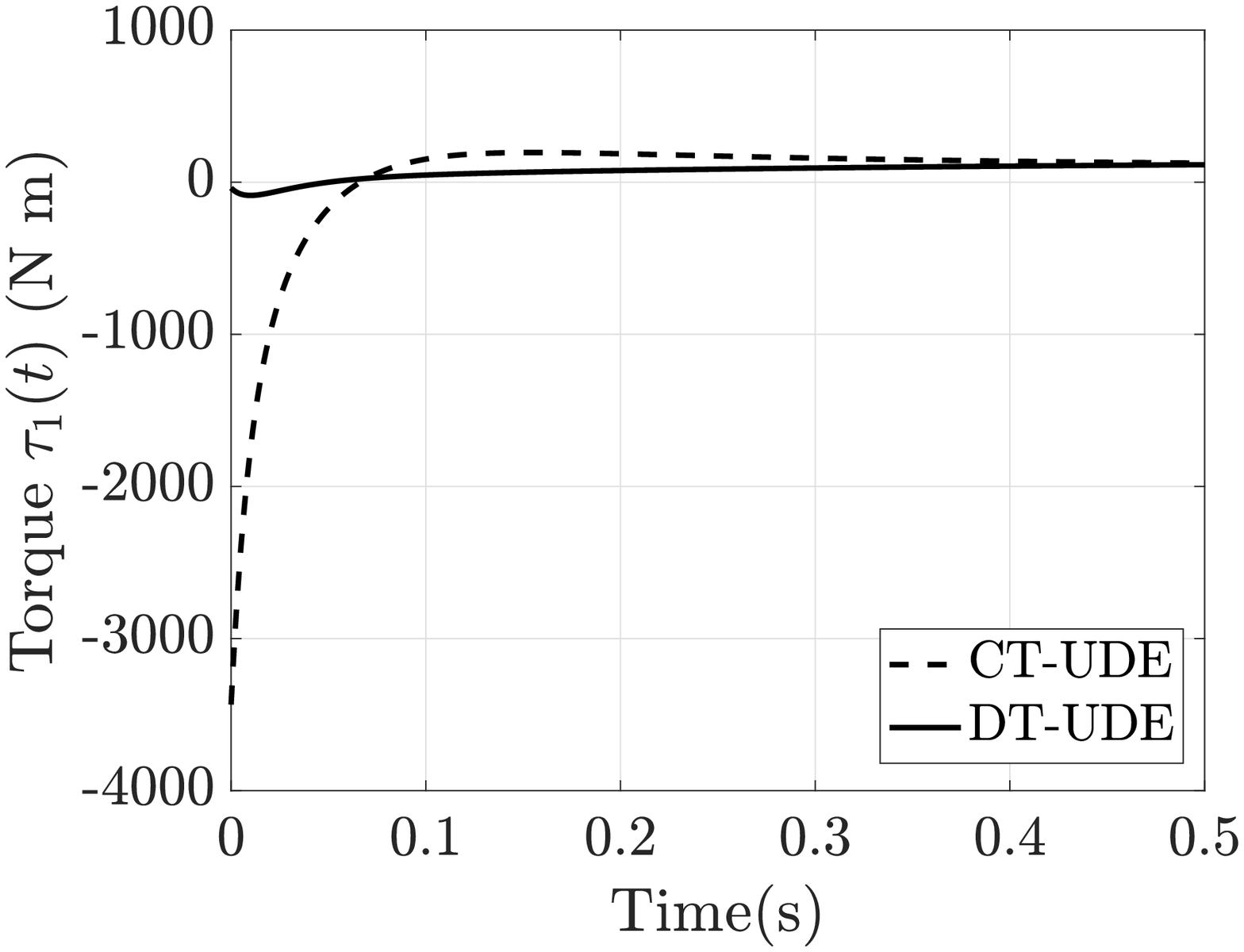}
	\caption{Torque $\tau_1(t)$}
\end{subfigure} \hspace{2cm}
\begin{subfigure}{0.4\textwidth}
	\centering
	\includegraphics[width = \textwidth]{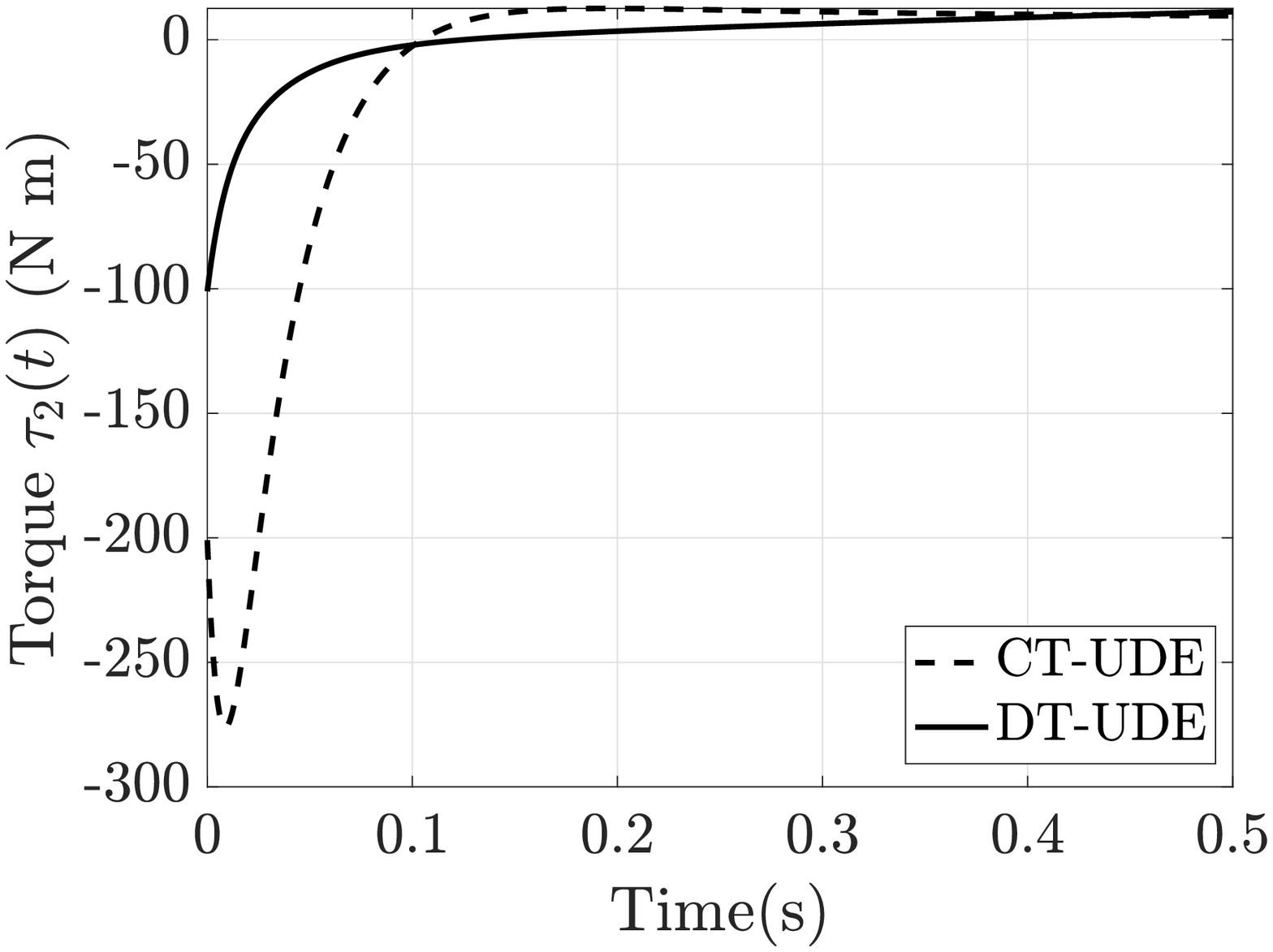}
	\caption{Torque $\tau_2(t)$}
\end{subfigure}

\caption{Comparing continuous-time and discrete-time UDE for non-zero initial conditions.}
\label{fig:CT-DT}
\hrulefill
\vspace{-0.3cm}

\end{figure*}

The results for this comparative study are shown in Fig. \ref{fig:Comparison}, in terms of output tracking error. It is evident that Designs $3$ and $4$, both based on UDE, outperform all other designs in terms of tracking performance. The performance specifications are not achieved using either gravity-compensated PD-control, or SMC. It now remains to be shown that the DT-UDE-based strategy can outperform continuous-time UDE. Note that throughout the simulations, zero initial conditions are assumed. This restriction is now relaxed, and initial conditions $\theta_1(0) = 1 \text{ rad}$, $\dot{\theta}_1(0) = 0 \text{ rad s}^{-1}$, $\theta_2(0) = -1 \text{ rad}$ and $\dot{\theta}_2(0) = 1 \text{ rad s}^{-1}$ are chosen. The eigenvalues of the closed-loop matrix is chosen to coincide with the eigenvalues of the reference model, in continuous-time and discrete-time as appropriate. It was illustrated in \cite{DT-UDE} that the discrete-time UDE strategy is less sensitive to initial values of tracking error, compared to continuous-time UDE, in terms of improved control energy. This is illustrated for the robot manipulator system in Fig. \ref{fig:CT-DT}. While the tracking error for $\theta_1(t)$ in Fig. \ref{fig:CT-DT}(a) converges slower for DT-UDE, convergence is marginally faster for $\theta_2(t)$. More significantly, the input torques $\tau_1(t)$ and $\tau_2(t)$ have much lower values for DT-UDE, as shown on an expanded time scale in Figs. \ref{fig:CT-DT}(c) and \ref{fig:CT-DT}(d). The peak value of $\tau_1(t)$ for DT-UDE is $0.04$ times the peak for continuous-time UDE, and the same ratio is around $1/3$ for $\tau_2(t)$.

To summarize, this section has demonstrated that the DT-UDE-based controller--observer structure performs significantly better than other well-known design strategies for the control of robot manipulators, resulting in improved tracking performance as well as lower control energy. This also subsequently indicates improved disturbance rejection compared to existing strategies. Further, the observer is used for estimating states for discrete-time UDE, whereas the actual plant states are used for feedback in all other strategies. The performance of other strategies is likely to degrade further with an observer in the loop.

\section{Concluding Remarks} \label{sec:Conclusion}
In this article, a robust discrete-time controller--observer structure based on the framework of Uncertainty and Disturbance Estimator (UDE) is presented. This addresses a drawback of the discrete-time UDE formulation in \cite{DT-UDE}, and only system outputs are required for control design in this work. The observer dynamics incorporate the estimate of the `lumped' disturbance acting on the system, in order to mimic the plant. The control law uses an auxiliary error based on state estimates, instead of using the state tracking error. The discrete-time observer in conjunction with the UDE-based controller result in a DT-UDE-based controller--observer structure. A detailed stability analysis is provided, based on both qualitative and quantitative methods. The entire strategy is simulated for the two-link robot manipulator system, and is shown to achieve highly accurate tracking performance, with excellent disturbance rejection. Further, the proposed strategy outperforms well-known techniques for the control of robot manipulators, including Sliding Mode Control (SMC), gravity-compensated proportional-derivative (PD) control and continuous-time UDE, in terms of improved tracking performance and lower control energy. An interesting avenue for future work is to modify the filter $G(\gamma)$ to achieve improved disturbance rejection. This can be accomplished using either higher-order filters or modification of $G(\gamma)$ by incorporating a parameter $\alpha$, and has been explored in \cite{IOL-UDE, Alpha} for continuous-time UDE.

\small{
\bibliographystyle{IEEEtran}
\bibliography{refs}
}

\end{document}